\documentclass[pra,twocolumn,preprintnumbers,amsmath,amssymb,superscriptaddress,showpacs,longbibliography]{revtex4-2}

\RequirePackage{fix-cm}
\RequirePackage{amsmath}

\usepackage[T1]{fontenc} 
\usepackage[utf8]{inputenc} 

\usepackage{graphicx}
\usepackage{latexsym}
\usepackage{amsmath}
\usepackage{graphics}
\usepackage{amssymb}
\usepackage{layout}
\usepackage{verbatim}
\usepackage{amsfonts,epsfig}
\usepackage{soul}
\usepackage[dvipsnames,table]{xcolor} 
\usepackage{hyperref}
\usepackage{tikz}
\usepackage[compat=1.1.0]{tikz-feynman}
\usepackage{color}
\hypersetup{colorlinks, linkcolor={blue!90!black}, citecolor={blue!90!black}, urlcolor={blue!90!black} }
\usetikzlibrary{quantikz2}

\newcommand{\beq}{\begin{equation}}
\newcommand{\eeq}{\end{equation}}
\newcommand{\bea}{\begin{eqnarray}}
\newcommand{\eea}{\end{eqnarray}}

\newcommand{\tr}{\hbox{Tr}}

\usepackage{color}

\usepackage{ulem} 

\renewcommand{\selectlanguage}[1]{}

\begin{document}

\title{Qubit-environment entanglement in time-dependent pure dephasing
}

\author{Ma{\l}gorzata Strza{\l}ka}
\affiliation{Institute of Physics (FZU), Czech Academy of Sciences, Na Slovance 2, 182 00 Prague, Czech Republic}
\affiliation{Faculty  of  Mathematics  and  Physics, Charles University, Ke Karlovu 5,
121 16 Prague, Czech Republic}
\author{Radim Filip}
\affiliation{
Department of Optics, Palack{\'y} University, 17. Listopadu 12, 771 46 Olomouc, Czech Republic}
\author{Katarzyna Roszak}
\affiliation{Institute of Physics (FZU), Czech Academy of Sciences, Na Slovance 2, 182 00 Prague, Czech Republic}

\date{\today}

\begin{abstract}
We show that the methods for quantification of system-environment entanglement that were recently
developed for interactions that lead to pure decoherence of the system can be straightforwardly
generalized to time-dependent Hamiltonians of the same type. This includes the if-and-only-if criteria of separability, as well as the entanglement measure applicable to qubit systems, and methods
of detection of entanglement by operations and measurements performed solely on the system
without accessing the environment. We use these methods to study the nature of the decoherence of a qubit-oscillator
system. Qubit-oscillator entanglement is essential for developing bosonic quantum technology with quantum non-Gaussian states and its applications in quantum sensing and computing. The dominating bosonic platforms, trapped ions, electromechanics, and superconducting circuits, are based on the time-dependent gates that use such entanglement to achieve new quantum sensors and quantum error correction.  The step-like time-dependence of the Hamiltonian that is taken
into account allows us to capture complex interplay between the build-up of classical
and quantum correlations, which could not be replicated in time-independent scenarios.
\end{abstract}
\maketitle

\section{Introduction \label{sec1}}

The study of the generation of entanglement between a system and its environment is typically hard,
because of the size and limited accessibility of the environment. 
The environment can be complex, but a single oscillator accessible only through a single qubit behaves as an environment, too. 
Although methods to calculate entanglement directly from 
a density matrix are available for two qubits \cite{wootters98,seidelmann22,schmolke22,vovk22}, when either of the systems becomes large
proper quantification of entanglement requires many-parameter-optimization \cite{vedral97,rains99}. The only 
available direct tool is Negativity \cite{vidal02,plenio05b}, but being based on the positive-partial-transpose (PPT)
criterion \cite{peres96a,horodecki96}, it does not capture bound entangled states \cite{horodecki98,smolin01,diguglielmo11,hiesmayr13,sentis18,gabodulin19} and is still numerically demanding
for larger systems. 

Recently, large progress has been made which allows efficient study of entanglement generation
for a class of system-environment Hamiltonians that lead to pure decoherence of the system,
as long as the initial state is of product form and the system of interest is in a pure state
(the state of the environment can be arbitrary). The pure dephasing types of interactions are essential in current quantum technology with superconducting circuits \cite{eickbusch22}, trapped ions \cite{katz23}, electromechanical oscillators \cite{ma21}, and, for a long time, in the cavity QED \cite{haroche20}. 

Firstly, the quantification of a system-environment state at a given time can now be qualified as 
separable or entangled can be performed with relative ease \cite{roszak15,roszak18}
and it has been shown that this type of interactions can lead to two distinct types of entanglement
for larger systems \cite{roszak18}. Furthermore, understanding of the nature of the correlations
that can be formed during the evolution, allowed the design of schemes that detect
this type of entanglement that are operated solely on the system of interest with no need
to access the environment \cite{roszak19a,rzepkowski20,strzalka21,roszak21}. The schemes work, because 
the build-up of entanglement leaves a distinct trace on the state of the environment
(which is related to the equivalence of this entanglement with quantum discord 
from the point of view of the environment \cite{roszak17}), which can
in turn affect the system evolution. 
The ease with which such entanglement can be detected
suggests that any quantum algorithm operated in a noisy setting will react differently to decoherence
of quantum and of classical origins. This has already been shown on the simplest algorithms, such
as teleportation \cite{harlender22,roszak23} and the spin echo \cite{roszak21}.

Beyond being resource for quantum technology, this type of Hamiltonian describes the most fundamental type of decoherence \cite{zurek03} that is not accompanied
by energy exchange between the system and the environment. It has been 
widely used in fundamental studies
of the nature of decoherence, and forms the basis for quantum Darwinism studies
and investigation of the nature of the quantum-to-classical transition today \cite{ollivier04,ollivier05,zurek09,korbicz14,horodecki15,mironowicz17,lampo17,roszak19b,lorenzo20,roszak20a,baldijao21}.
Furthermore, this type of decoherence tends to dominate in realistic solid state systems where the 
energy of environment quanta is much smaller than the energy level separation in the system
of interest,
such as excitonic and electronic states confined in quantum dots \cite{borri01,vagov03,vagov04,glassl13,tahara14,salamon17,seidelmann19} and various types of 
spin qubits \cite{zhao12,kwiatkowski18,bartling22,bayliss22,wang22,onizhuk23}. 

In this paper we show that the methods previously devised for pure decoherence 
can be generalized to time-dependent Hamiltonians of the same type, because time-dependence
does not change the nature of the correlations that can be generated in any fundamental way. 
This includes the if-and-only-if criteria of separability for a single qubit system \cite{roszak15},
and for a system of any size \cite{roszak18}, as well as the single-qubit entanglement measure
\cite{roszak20}. Also all of the schemes for entanglement detection \cite{roszak19a,rzepkowski20,strzalka21,roszak21}, which are a direct consequence
of the form of the separability criteria can be used in case of a time-dependent interaction.

As described above, it is of special interest 
for the study of hybrid solid-state-optical systems, such as superconducting
transmon qubits \cite{touzard19,campagne20,gao21,delaney22,hassani23} and trapped 
ions \cite{leibfried03,kienzler16,lv18,bock18,landsman19,monroe21,kokail22,matsos23}. The biggest difference with respect to standard solid-state qubits here, is the possibility
of engineering and control of the interaction with the optical environment. This means that 
for such systems, the interaction can be specially tailored to control the level of entanglement
build up between the system and its environment in order to be used as a resource \cite{chaves10,chitambar19}, e.~g.~for
decoherence control \cite{roszak15b,rzepkowski23,roszak23}, and hence a deeper understanding
of such entanglement is critical.

We use the time-dependent methods for the study of qubit-environment entanglement
generated via a tunable Hamiltonian which describes the interaction with a single
bosonic mode that is used both for the description 
of a transmon qubit interacting with microwave cavity photons \cite{touzard19,campagne20}
as well as trapped ions interacting with mechanical oscillator modes \cite{monroe21}, or electrically controlled mechanical modes \cite{ma21}. 
For these systems, the interaction
leads to pure decoherence and it is experimentally controllable to a high extent.   
We change the interaction in a step-like manner between such that does not lead to entanglement
generation of the qubit with an initial mixture of Fock states, and such that does. 
Both the non-entangling and entangling interactions lead to decoherence of the qubit,
but regardless of the similarities in the qubit evolution for the two interactions studied 
separately, their nature is very different. This is visible when the interactions act consecutively
on the qubit, which leads to the build-up of quantum correlations affecting entanglement-driven
decoherence, and vice versa. We observe nontrivial effects such as the simultaneous growth of 
coherence and entanglement at certain time periods, as well as the non-entangling Hamiltonian
driving entanglement when it is preceded by a time when the interaction is entangling. 
These effects could not be observed using the time-independent methods, even though
the evolution of the density matrix is obtained in a series of time-independent steps,
because any correlations (quantum or classical) formed between the system and the environment
preclude their application.

The article is organized as follows. In Sec.~\ref{sec2} we introduce the time-dependent pure 
dephasing Hamiltonian and describe the formalism to obtain the resulting time-evolution that we
use in the rest of the paper. In Sec.~\ref{sec3} we show how the time-independent separability criteria
for system-environment entanglement are generalized to the time-dependent scenario.
This holds true also for the qubit-environment entanglement measure and we use it is Sec.~\ref{sec3a} 
to study the evolution of entanglement between a transmon/trapped-ion qubit and its environment.
Sec.~\ref{sec5} concludes
the paper.

\section{The Hamiltonian and the evolution \label{sec2}}

\begin{figure}[!tb]
	\begin{quantikz}
		\lstick[ label style={label={[blue]below:qubit system}}]{$\sum_ic_i\ket{i}$}&\ctrl{1}&\gate[2,style={dashed,rounded
			corners,fill=blue!20}]{\hat{\sigma}(t)} &  \\
		\lstick[label style={label={[blue]below:oscillator environment}}]{$\hat{R}(0)$} &\gate{\hat{w}_{ii}(t)} & &\\
	\end{quantikz}
	\caption{Circuit representing entanglement generation during pure decoherence.
		The system state is initially in a superposition of pointer states $(\sum_ic_i\ket{i})$ and the environment
		is in an arbitrary, possibly mixed state $\hat{R}(0)$. The interaction acts as a gate $\hat{w}_{ii}(t)$ on the
		environment which is conditional on the pointer state of the system, $\ket{i}$, yielding the system-environment
		state at time $t$, $\hat{\sigma}(t)$, given by eq.~(\ref{sigmag}).
		\label{g0}}
\end{figure}
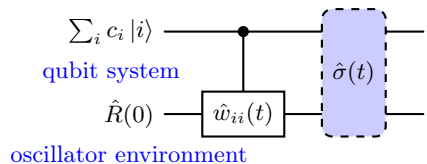

We are interested time-dependent Hamiltonians that describe system-environment (SE) evolution which 
leads to pure decoherence (PD) of the 
system when the degrees of freedom of the environment are traced out.
To this end we must first specify the general form of such Hamiltonians.
The general conditions for PD evolutions hold regardless of time-dependence, namely
that the free system Hamiltonian must commute with the interaction term, but now
time-dependence imposes that this condition must be fulfilled at all times $t$ and $t'$,
\begin{equation}
\label{comutg}
\left[\hat{H}_S(t),\hat{H}_{int}(t')\right]=0.
\end{equation}
Here we assumed that the full SE Hamiltonian is of the form 
$\hat{H}_{PD}(t)=\hat{H}_S(t)+\hat{H}_E(t)+\hat{H}_{int}(t)$, where the first two terms on the right
describe the free Hamiltonians of the system and the environment, respectively, while
the third term describes their interaction.

The commutation relation (\ref{comutg}) translates into limitations on the possible forms
of the system and iteration Hamiltonian. Most importantly their fulfillment requires there to 
exist a well defined and time-independent pointer basis of the system, which we will denote as
$\{|i\rangle\}$, with $i=0,\dots ,N-1$, where $N$ is the dimension of the system.
Hence the time-dependence of the system Hamiltonian has to be limited to the eigenvalues,
while the time-dependence of the interaction is fully described by environmental operators.
We can now explicitly write the most general form of a time-dependent PD Hamiltonian,
\begin{equation}
\label{hampdt}
\hat{H}_{PD}(t)=\sum_i|i\rangle\langle i|\otimes\hat{V}_i(t),
\end{equation}
where only the environmental operators $\hat{V}_i(t)$ are time-dependent.
They describe contributions to the Hamiltonian from all three terms and can be written as
\begin{equation}
\label{vi}
\hat{V}_i(t)=\varepsilon_i(t)+\hat{H}_E(t)+\tilde{V}_i(t),
\end{equation}
where $\varepsilon_i(t)$ is the eigenvalue of $\hat{H}_S(t)$ corresponding to pointer state $|i\rangle$
and $\tilde{V}_i(t)$ are environmental operators which describe the effect of a given system
pointer state on the environment, obtained by writing the interaction Hamiltonian
in the form 
\begin{equation}
\label{vint}
\hat{H}_{int}(t)=\sum_i|i\rangle\langle i|\otimes\tilde{V}_i(t).
\end{equation}

Once the general form of PD Hamiltonians is specified, one can easily find the form of the 
evolution operator, which is analogous to the time-independent PD evolution operator \cite{roszak15,roszak18},
\begin{equation}
\label{ug}
\hat{U}_{PD}(t)=\sum_i|i\rangle\langle i|\otimes\hat{w}_i(t).
\end{equation}
The critical difference here lies in the form of the conditional evolution operators of the
environment $\hat{w}_i(t)$ which are given by
\begin{equation}
\label{wig}
\hat{w}_i(t)=\mathrm{Texp}\left[-\frac{i}{\hbar}\int_0^t dt'\hat{V}_i(t')\right],
\end{equation}
where $\mathrm{Texp}\left[\dots\right]$ denotes the time-ordered exponential function.
It is important to note here that although the operators (\ref{wig}) can have a much more
complicated structure than their time-independent counterparts, they are still
unitary operators.

Having found the evolution operator, one can write the SE density matrix at time $t$
for any initial conditions. For a product initial SE state with a pure system state
given by $|\psi(0)\rangle=\sum_i c_i |i\rangle$ and an arbitrary initial state of the environment
described by the density matrix $\hat{R}(0)$, this is given by
\begin{equation}
\label{sigmag}
\hat{\sigma}(t)=\sum_{ij}c_ic_j^*|i\rangle\langle j|\otimes\hat{R}_{ij}(t),
\end{equation}
with
\begin{equation}
\label{rijg}
\hat{R}_{ij}(t)=\hat{w}_i(t)\hat{R}(0)\hat{w}_j^{\dagger}(t).
\end{equation}
The effect of the interaction, which can be interpreted as a conditional gate, where the evolution of the
environment is conditional on the pointer state of the qubit, is illustrated in Fig.~\ref{g0}.

\section{System-environment entanglement \label{sec3}}

The SE density matrix (\ref{sigmag}) has exactly the same structure as the one studied in Ref.~\cite{roszak18} in order to qualify SE states obtained during time-independent PD evolution
as entangled or separable. This structure together with the fact that the conditional
evolution operators of the environment (\ref{wig}) are unitary, allows us to directly transcribe
the complete set of separability conditions from time-independent PD Hamiltonians
to the time-dependent case.
The proofs from Ref.~\cite{roszak15} for
qubit-environment entanglement (QEE), and the generalized proofs for system-environment 
entanglement (SEE) of Ref.~\cite{roszak18} hold for time-dependent PD described by the 
Hamiltonian (\ref{hampdt}) as long as the initial SE state is of product form with a pure initial
state of the system [which is required to obtain eq.~(\ref{sigmag})].

If the system under study is a qubit, there exists a unique separability criterion
for pure dephasing \cite{roszak15}, namely a qubit is separable from its environment
if and only if 
\begin{equation}
\label{crit0}
\hat{R}_{00}(t)=\hat{R}_{11}(t),
\end{equation}
where $\hat{R}_{00}(t)$ and $\hat{R}_{11}(t)$ are given by eq.~(\ref{rijg}). Otherwise there
is QEE in the system.
This makes checking for QEE particularly straightforward and 
allowed for the existence of an entanglement measure which can be calculated directly from a density matrix
obtained during PD evolution \cite{roszak20}.
This measure is also valid for time-dependent PD Hamiltonians and is given by
\begin{equation}
\label{meas}
E(t)=4|c_0|^2|c_1|^2\left[1-F\left(\hat{R}_{00}(t),\hat{R}_{11}(t)\right)\right],
\end{equation}
where $F\left(\hat{\rho}_{1},\hat{\rho}_{2}\right)=
\left[\tr\sqrt{\sqrt{\rho_1}\rho_2\sqrt{\rho_1}}\right]^2$
is the Fidelity. Hence the amount of entanglement that forms between a qubit and its environment
during PD depends on the initial qubit coherence, $|\rho_{01}(0)|^2=|c_0|^2|c_1|^2$,
and evolves proportionally to how different the state of the environment becomes for the two qubit
pointer states. 

For larger systems, there are two types of separability criteria and a system of size $N$,
there exist $N-1$ independent criteria of the first type and $(N-1)(N-2)/2$ independent
criteria of the second type \cite{roszak18}. 
If any one of the following criteria is broken at time $t$ this means that 
there is entanglement between the system and the environment at this time.

Separability criteria of the first type state that for all $i\neq j$ we must have
\begin{equation}
\label{crit1}
\hat{R}_{ii}(t)=\hat{R}_{jj}(t),
\end{equation}
meaning that 
at a given time 
the state of the environment under the condition that the qubit is in pointer state
$|i\rangle$ is the same as its state when the qubit is in state $|j\rangle$.
The QEE criterion is a separability criterion of this type. 

Criteria of the second type are more abstract in interpretation and relate to commutation between 
pairs of conditional evolution operators. Namely, they state that for all 
for all $i,j,k,l$ we must have 
\begin{equation}
\label{crit2}
\left[\hat{w}_i(t)\hat{w}_j^{\dagger}(t),\hat{w}_k(t)\hat{w}_l^{\dagger}(t)\right]=0
\end{equation}
for separability.
These criteria are related to internal SE coherences and criteria of this type do not exist
if the system under study is a qubit.

Incidentally, conditions of the first type (\ref{crit1}) cannot be broken when the initial density matrix
of the environment is a fully mixed state, but conditions of the second type (\ref{crit2}) can.
This means that if a system is a qubit (so there are no conditions of the second type),
PD interactions cannot lead to entanglement with a maximally mixed environment, but for larger systems it is possible.
This was shown for a qutrit system in an example in Ref.~\cite{roszak18}.

Entanglement which is accompanied by the violation of any criterion 
of the first type (\ref{crit1}) can be detected experimentally, because it manifests itself
directly in the state of the environment. One could measure observables on the environment
to witness SEE, but such measurements are hard in general (with the actual experimental feasibility
depending strongly on the physical system under study). Yet, for time-independent PD Hamiltonians,
it has been shown that the
operation of simple algorithms on the system without the need to access the environment
are sufficient for the detection of QEE in many situations for qubits \cite{roszak19a,rzepkowski20} and for larger systems \cite{strzalka21}.
Since there is no qualitative change in the SE density matrix (\ref{sigmag}) which is obtained as a result
of a time-dependent Hamiltonian, the schemes introduced in Refs \cite{roszak19a,rzepkowski20,strzalka21} can be operated as entanglement witnesses also in time-dependent scenarios.

\section{Transmon qubit and microwave cavity/ Trapped ion and mechanical oscillator mode  \label{sec3a}}

As an example we will study the evolution of entanglement using the measure (\ref{meas}) 
for an interaction Hamiltonian which can describe the effective coupling of a superconducting transmon qubit to the microwave cavity modes, as well as the interaction between a qubit defined on a trapped ion 
	and environment of a mechanical oscillator mode. The Hamiltonian is
given by \cite{touzard19,campagne20,katz23}
\begin{equation}
\label{ham}
\hat{H}(t)=\hat{\sigma}_z\otimes\left[\left(\alpha(t)\hat{a}^{\dagger}+\alpha^*(t)\hat{a}\right)+\beta\hat{a}^{\dagger}\hat{a}
+\gamma(t)\right].
\end{equation}
Here, $\hat{\sigma}_z$ is the appropriate Pauli operator acting on the qubit subspace, while operators $\hat{a}^{\dagger}$ and $\hat{a}$
are creation and annihilation operators in the subspace of the environment.
Time dependence is explicitly marked when applicable 
and $\gamma(t)$ is responsible for free evolution of the qubit.

The Hamiltonian (\ref{ham}) can be easily rewritten into the pure-dephasing (PD) form
given by equation (\ref{hampdt}), with $i=0,1$ and
\begin{equation}
\label{v}
\hat{V}_{0/1}(t)=\pm\left[\left(\alpha(t)\hat{a}^{\dagger}+\alpha^*(t)\hat{a}\right)+\beta\hat{a}^{\dagger}\hat{a}
+\gamma(t)\right].
\end{equation}
Since the environmental operators $\hat{V}_{0/1}(t)$ not only commute, but differ only by the sign,
it is easy to show that
the functions $\hat{w}_{0/1}(t)$ commute at any given time,
\begin{equation}
\label{comut}
\forall_{t}\left[\hat{w}_0(t),\hat{w}_1(t)\right]=0,
\end{equation}
since $\hat{w}_0(t)=\hat{w}_1^{\dagger}(t)$.
This does not translate however into them commuting at different times as it would in time-independent
cases.

In the following, we will be considering the simplest case in terms of the time-dependence 
of the Hamiltonian (\ref{ham}), namely such that 
the parameter $\alpha(t)$ is a step function.
We assume that initially $\alpha(t)=0$ until time $t_1$, then it is constant $\alpha(t)=\alpha$
for duration $t_2$, and again $\alpha(t)=0$ for a time duration $t_3$,
\begin{equation}
\label{alpha}
\alpha(t)=\left\{\begin{array}{cc}
	0&\mathrm{for}\; t\in[0,t_1)\\
	\alpha&\mathrm{for}\; t\in[t_1,t_1+t_2)\\
	0&\mathrm{for}\; t\in[t_1+t_2,t_1+t_2+t_3].
	\end{array}\right.
\end{equation}
The actual value of $\gamma(t)$ is irrelevant, since it does not influence
the generated entanglement, nor the evolution of the degree of qubit coherence 
(absolute value of the off-diagonal element of the density matrix).
We choose the step-function time-dependence of the Hamiltonian, because it allows us to observe
behaviors of the time evolution of entanglement which are not possible for time-independent Hamiltonians,
while it's simplicity allows for a straightforward interpretation of the observed results in terms 
of the generation of different types of correlations between the qubit and the environment.

For this scenario the conditional evolution operators of the environment $\hat{w}_k(t)$, with $k=0,1$,
consist of three parts
\begin{equation}
    \hat{w}_k(t) 
         = \hat{w}_k^3(t_3)\hat{w}_k^2(t_2)\hat{w}_k^1(t_1),
\end{equation}
where the operators $\hat{w}_k^i(t_i)$ are given by
\begin{subequations}
 \begin{eqnarray}
      \hat{w}_{0/1}^{1/3}(t) 
        &=&e^{\mp\frac{i}{\hbar}\beta\hat{a}^\dagger\hat{a}t},\\
        \hat{w}_{0/1}^{2}(t)  &=&e^{Y_{0/1}(t)}e^{i\Phi_{0/1}(t)}e^{\mp\frac{i}{\hbar}\beta\hat{a}^\dagger\hat{a}t},
\end{eqnarray}
\end{subequations}
with
\begin{eqnarray}
\nonumber
\Phi_{0/1}(t)&=&\mp\frac{|\alpha|^2}{\beta^2}\sin{\frac{\beta t}{\hbar}},\\
\nonumber
Y_{0/1}(t)&=&\frac{\alpha}{\beta}(e^{\mp\frac{i}{\hbar}\beta t}-1)\hat{a}^\dagger-\frac{\alpha^*}{\beta}(e^{\pm\frac{i}{\hbar}\beta t}-1)\hat{a}.
\end{eqnarray}

We will consider two types of initial states for the environment,
while the qubit will always initially be in an equal superposition state.
Firstly, the environment will be initially at a thermal equilibrium of Fock states,
meaning the Gibbs state corresponding to the Hamiltonian $\hat{H}_0=\Gamma\hat{a}^{\dagger}\hat{a}$,
and later we will show plots for coherent states for comparison.

\begin{figure}[!tb]
\centering
	\includegraphics[width=1.\columnwidth]{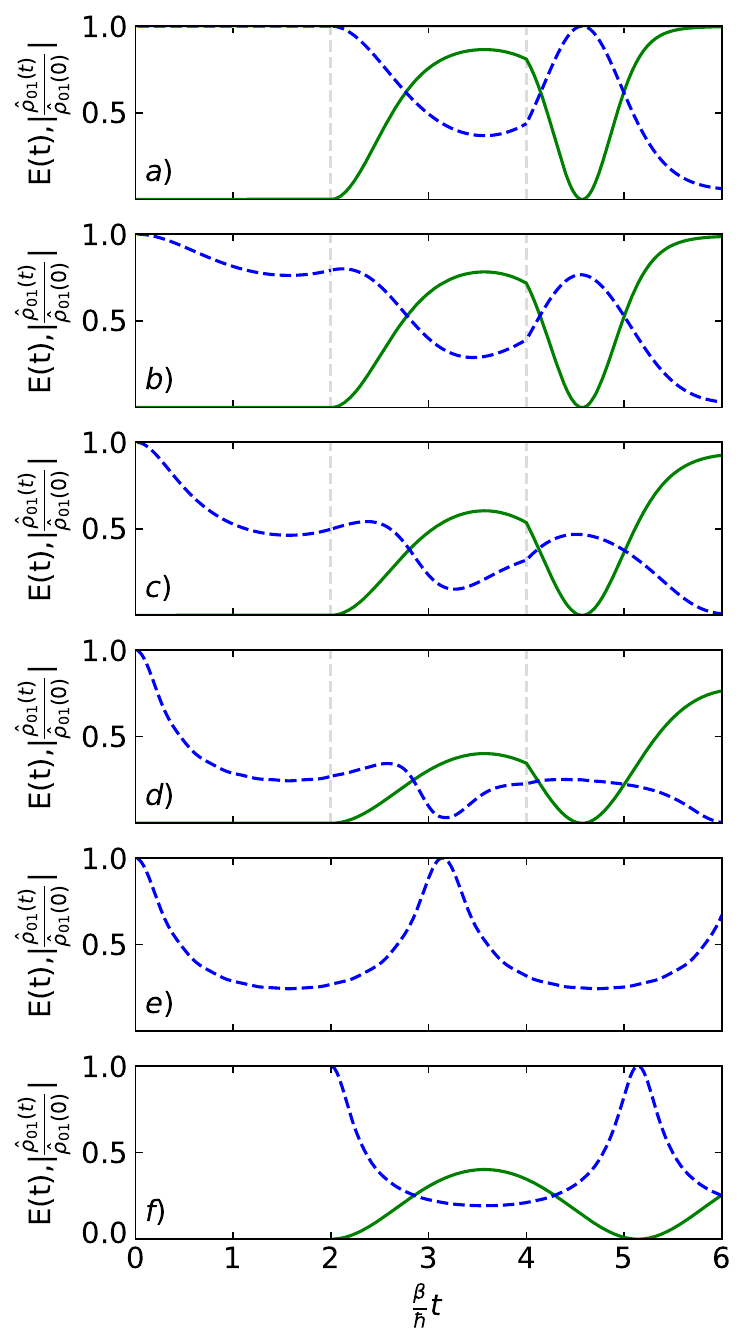}
	\caption{Evolution of QEE (solid lines) and normalized qubit coherence (dashed lines) 
		for initial Gibbs state of the environment at different temperatures: (a) $k_BT/\Gamma =0$, (b) $k_BT/\Gamma =0.5$, (c) $k_BT/\Gamma =1$, (d) $k_BT/\Gamma =2$. 
		Vertical dashed lines denote the times when parameter $\alpha$ is changed
		from $\alpha=0$ to $\alpha\neq 0$ at $\hbar t/\beta = 2$ and back from $\alpha\neq 0$ to
		$\alpha=0$ at $\hbar t/\beta = 4$.
		The Hamiltonian parameters are set to $\alpha/\beta=(1+i)/2$ for $\alpha\neq 0$.
		(e) Evolution with constant Hamiltonian and $\alpha=0$ for $k_BT/\Gamma =2$.
		(f) Evolution with constant Hamiltonian with $\alpha/\beta=(1+i)/2$ for $k_BT/\Gamma =2$.
		The plot is shifted for easier comparison and the evolution starts at $\hbar t/\beta = 2$.
}\label{fig2}
\end{figure}

In Fig.~\ref{fig2} QEE measured by eq.~(\ref{meas}) is plotted by the solid lines 
as a function of time. Complementarily, the absolute value of the qubit coherence normalized 
by it's initial value, $\frac{\hat{\rho}_{01}(t)}{\hat{\rho}_{01}(t)}$, is plotted using dashed lines. 
Plots (a-d) contain the evolution for the three-step time-dependence of Hamiltonian (\ref{ham})
with the parameter $\alpha$ changing as given by eq.~(\ref{alpha}) at $\beta t/\hbar=\beta t_1/\hbar =2$ and 
$\beta t/\hbar=\beta (t_1+t_2)/\hbar =4$ which are marked
by gray vertical lines on the plots. 
The upper panel (a) contains zero-temperature results, while progressively higher temperatures are
taken into account in the lower plots, (b) $k_BT/\Gamma =0.5$, (c) $k_BT/\Gamma =1$, (d) $k_BT/\Gamma =2$. 

Let us first note that there are no qualitative changes in the evolution of entanglement when the
temperature is increased, but there is a stark difference in the decoherence before time $t_1$ is 
reached at zero temperature (which is the only situation when the first part of the evolution
does not display decoherence). This is because for $t\le t_1$ decoherence is not an outcome
of the generation of entanglement between the qubit and it's environment, but rather the establishment
of classical correlations between them. For pure states, classical correlations cannot be generated
through a unitary evolution and thus decoherence is not possible at zero temperature.

Entanglement starts being generated after $\beta t/\hbar=2$ because terms of the Hamiltonian (\ref{ham}) with $\alpha\neq 0$ which are responsible for the conditional evolution of the environment do not commute
with $\hat{R}_{00}(t_1)=\hat{R}_{11}(t_1)$. After time $\beta t/\hbar=4$ when $\alpha$ is again set to zero,
the qubit-environment interaction is nevertheless capable of driving the evolution of entanglement,
because of the QE correlations that have been established in the previous phase of the evolution.
Note that the third part of evolution is different both for entanglement and coherence, which
manifests itself most visibly in the sharp change observed at $\beta t/\hbar=4$.

For comparison, we have additionally plotted the evolution of entanglement and coherence for the same
Hamiltonian, but without time-dependence, with $\alpha=0$ in Fig.~\ref{fig2} e)
and with $\alpha\neq 0$ in Fig.~\ref{fig2} f) (here the evolution starts at $\beta t/\hbar=2$
in order to ease the comparison between these curves and analogous evolution that has been 
preceded by an interaction with $\alpha=0$).
The plots correspond to $k_BT/\Gamma =2$ as in Fig.~\ref{fig2} d).
The oscillatory behavior observed in panels e) and f) would also be present in the time-dependent
evolution (a-d) if the transition times between different values of $\alpha$ were chosen longer,
as this is a trivial consequence of only one bosonic mode being taken into account. Comparison 
of panels e) and f) shows that there is no qualitative change in the evolution of coherence,
even though the nature of the decoherence is fundamentally different, as one is the result of
classical SE correlations being established, while the other is driven by entanglement generation.
Quite surprisingly, when the switch between the two types of interactions is made in Fig.~\ref{fig2} d), we observe a stark qualitative change in decoherence, due to the 
interplay of classical and quantum correlations, even though the actual generation of entanglement
in panels d) and f) resemble each other closely.

Because of the time-dependence of the Hamiltonian (\ref{ham}) we are able to observe 
specific features of entanglement evolution which are otherwise rare. Firstly there is a transition
between decoherence classical in nature and decoherence which is induced by QEE, as described above,
but in several time instances we see that the qubit coherence can grow at the same time
as entanglement does. This is only possible at finite temperatures and is the outcome of the competition
between quantum and classical decoherence mechanisms. This is most distinct just after $\beta t/\hbar=2$
when the classical dephasing process leads to the enhancement of qubit coherence
due to the unitary nature of the QE evolution and the single bosonic mode taken into account,
while $\alpha(t)=const$ ensures the establishment of quantum correlations which start to lead to 
qubit decoherence.

It is important to note here that although the way that the time-dependence of the Hamiltonian
is taken into account
allows us to obtain the QE evolution by superposing time-independent evolution operators,
time-independent methods for the quantification/qualification of QEE would not be sufficient here.
This is because the SE states at $\beta t/\hbar=2$ and $\beta t/\hbar=4$ contain SE correlations (classical for $\beta t/\hbar=2$
and both quantum and classical for $\beta t/\hbar=4$)
and do not fulfill the requirements for initial SE states in the time-independent methods. 

\begin{figure}[!tb]
	\centering
	\includegraphics[width=1.\columnwidth]{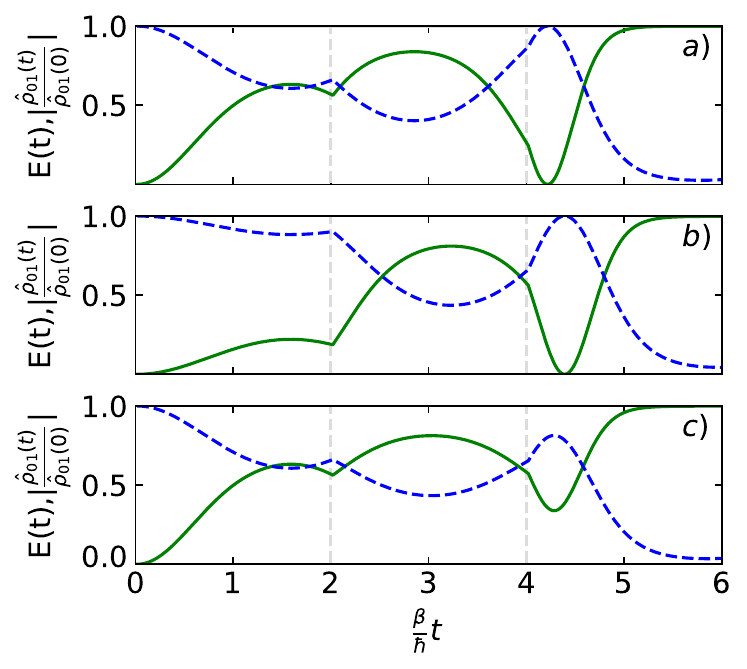}
	\caption{Evolution of QEE (solid lines) and normalized qubit coherence (dashed lines) 
		for initial coherent state of the environment with a) $\zeta=0.5e^{i\pi/4}$,  b) $\zeta=0.25e^{i\pi/4}$, and  c) $\zeta=0.5$.
		  }\label{fig3}
\end{figure}

For completeness in Fig.~\ref{fig3} we plot the evolution of entanglement and coherence analogous to 
the plots in Fig.~\ref{fig2} for the situation when the initial state of the environment is
a coherent state,
\begin{equation}
\label{cstate}
             \ket{\zeta} = e^{-\frac{1}{2}|\zeta|^2}e^{\zeta\hat{a}^\dagger}e^{-\zeta^*\hat{a}}\ket{0}
\end{equation}
where $\zeta$ is a complex number. The panels correspond to different values of $\zeta$, and 
it varies only in amplitude between panels a) and b) and only in phase between panels a) and c).
This yields to a stark difference between the observed curves in the first phase when $\alpha=0$
between panels a) and b) which diminishes in the later phases,
while between panels a) and c) the biggest difference in the evolution is in the third
phase, when the parameter $\alpha$ is again set to zero.

Nevertheless, the most important difference manifests itself in the comparison between Figs \ref{fig2}
and \ref{fig3}, since these differences are most distinctly qualitative. For coherent states,
$\alpha=0$ does not preclude the generation of SEE from the initial product state, so entanglement 
is generated throughout the evolution. This exemplifies that in the generation of entanglement, not
only the SE interaction is important, but its interplay with the initial part of the environment
plays a critical part. 

As a last remark in this section, it is relevant to note that although the type of time-dependence
that has been included in the example under study is as simple as possible, it is sufficient to 
demonstrate nontrivial properties of the evolution of entanglement. These results could not be obtained outside of the time-dependent 
formalism. Furthermore modeling time-dependence as a number of small, consecutive steps is 
a fairly standard procedure \cite{fluhmann19,ma21}, so in principle the same method could be used
to model any time-dependence in the Hamiltonian.

\section{Conclusion \label{sec5}}

We have shown that the whole array of methods developed for the qualification and quantification of
entanglement that can be generated during the joint evolution of a system and its environment
which leads to pure decoherence of the system
can be generalized to time-dependent pure-decoherence Hamiltonians.
This is because the SE density matrix obtained during such an evolution
is qualitatively the same as in the time-independent case and the nature of the correlations 
that can be formed does not change. 
Hence, the same criteria for the qualification of SE states as separable or mixed
can be used. 

Nevertheless, time-dependence in the Hamiltonian allows for a much more complex evolution of
entanglement, e.~g.~in the extreme case interactions that only lead to build up of classical correlations during 
decoherence can be interchanged with such that rely on entanglement generation. 
We demonstrate this studying an interaction Hamiltonian that is used to describe
a transmon qubit interacting with a microwave cavity as well as a trapped ion interacting 
with mechanical modes. Such systems are good examples of systems
that effectively undergo pure decoherence while the environment is engineered
and the parameters of the interaction can be experimentally manipulated.

We use a step like time-dependence of the Hamiltonian, changing from an interaction which is 
non-entangling for an initial thermal-equilibrium
state of the environment, to an entangling one, and back. This allows us to show the stark change
in both the nature and time-dependence of the decoherence and of entanglement. In the first part
of the evolution, decoherence is not an effect of entanglement generation, but of the formation of
classical correlations
between the system and the environment. Once the interaction is switched to entangling,
there is an interplay between classical and quantum correlations which is reflected in the
decoherence, that can now display counter-intuitive behaviors, such as the reversal of decoherence
while entanglement grows. In the third part, when the entangling part of the evolution is switched
off, we still observe entanglement evolution, and we show that entanglement can 
grow in this phase to higher levels than the maximum obtained while the entangling interaction was
turned on. This is again due to the interplay of quantum and classical correlations that are 
present in the system at the moment when the nature of the interaction is changed.

It is important to stress here that the nontrivial features of the presented results could not
be replicated without the use of time-dependent methods. The evolution of coherence for the entangling Hamiltonian
is very different if it is not preceded by a period of classical-correlation-driven decoherence. 
Similarly, the fact that entanglement evolves and can grow in the third part of the evolution,
would not be possible if it were
not preceded by a period of the evolution when SEE was generated.

\section*{Acknowledgment}
R.F. acknowledges grant No. 22-27431S of the Czech Science Foundation and the project 8C22001 (SPARQL) of MEYS Czech Republic and the funding from the European Union’s Horizon 2020 research and innovation framework program under Grant Agreement No. 731473 and 101017733. This project has received funding from the European Union’s 2020 research and innovation programme (CSA—Coordination and support action, H2020-WIDESPREAD-2020-5) under grant agreement No. 951737 (NONGAUSS).


\begin{thebibliography}{73}%
	\makeatletter
	\providecommand \@ifxundefined [1]{%
		\@ifx{#1\undefined}
	}%
	\providecommand \@ifnum [1]{%
		\ifnum #1\expandafter \@firstoftwo
		\else \expandafter \@secondoftwo
		\fi
	}%
	\providecommand \@ifx [1]{%
		\ifx #1\expandafter \@firstoftwo
		\else \expandafter \@secondoftwo
		\fi
	}%
	\providecommand \natexlab [1]{#1}%
	\providecommand \enquote  [1]{``#1''}%
	\providecommand \bibnamefont  [1]{#1}%
	\providecommand \bibfnamefont [1]{#1}%
	\providecommand \citenamefont [1]{#1}%
	\providecommand \href@noop [0]{\@secondoftwo}%
	\providecommand \href [0]{\begingroup \@sanitize@url \@href}%
	\providecommand \@href[1]{\@@startlink{#1}\@@href}%
	\providecommand \@@href[1]{\endgroup#1\@@endlink}%
	\providecommand \@sanitize@url [0]{\catcode `\\12\catcode `\$12\catcode
		`\&12\catcode `\#12\catcode `\^12\catcode `\_12\catcode `\%12\relax}%
	\providecommand \@@startlink[1]{}%
	\providecommand \@@endlink[0]{}%
	\providecommand \url  [0]{\begingroup\@sanitize@url \@url }%
	\providecommand \@url [1]{\endgroup\@href {#1}{\urlprefix }}%
	\providecommand \urlprefix  [0]{URL }%
	\providecommand \Eprint [0]{\href }%
	\providecommand \doibase [0]{https://doi.org/}%
	\providecommand \selectlanguage [0]{\@gobble}%
	\providecommand \bibinfo  [0]{\@secondoftwo}%
	\providecommand \bibfield  [0]{\@secondoftwo}%
	\providecommand \translation [1]{[#1]}%
	\providecommand \BibitemOpen [0]{}%
	\providecommand \bibitemStop [0]{}%
	\providecommand \bibitemNoStop [0]{.\EOS\space}%
	\providecommand \EOS [0]{\spacefactor3000\relax}%
	\providecommand \BibitemShut  [1]{\csname bibitem#1\endcsname}%
	\let\auto@bib@innerbib\@empty
	\bibitem [{\citenamefont {Wootters}(1998)}]{wootters98}%
	\BibitemOpen
	\bibfield  {author} {\bibinfo {author} {\bibfnamefont {W.~K.}\ \bibnamefont
			{Wootters}},\ }\bibfield  {title} {\bibinfo {title} {Entanglement of
			formation of an arbitrary state of two qubits},\ }\href
	{https://doi.org/10.1103/PhysRevLett.80.2245} {\bibfield  {journal} {\bibinfo
			{journal} {Phys. Rev. Lett.}\ }\textbf {\bibinfo {volume} {80}},\ \bibinfo
		{pages} {2245} (\bibinfo {year} {1998})}\BibitemShut {NoStop}%
	\bibitem [{\citenamefont {Seidelmann}\ \emph {et~al.}(2022)\citenamefont
		{Seidelmann}, \citenamefont {Schimpf}, \citenamefont {Bracht}, \citenamefont
		{Cosacchi}, \citenamefont {Vagov}, \citenamefont {Rastelli}, \citenamefont
		{Reiter},\ and\ \citenamefont {Axt}}]{seidelmann22}%
	\BibitemOpen
	\bibfield  {author} {\bibinfo {author} {\bibfnamefont {T.}~\bibnamefont
			{Seidelmann}}, \bibinfo {author} {\bibfnamefont {C.}~\bibnamefont {Schimpf}},
		\bibinfo {author} {\bibfnamefont {T.~K.}\ \bibnamefont {Bracht}}, \bibinfo
		{author} {\bibfnamefont {M.}~\bibnamefont {Cosacchi}}, \bibinfo {author}
		{\bibfnamefont {A.}~\bibnamefont {Vagov}}, \bibinfo {author} {\bibfnamefont
			{A.}~\bibnamefont {Rastelli}}, \bibinfo {author} {\bibfnamefont {D.~E.}\
			\bibnamefont {Reiter}},\ and\ \bibinfo {author} {\bibfnamefont {V.~M.}\
			\bibnamefont {Axt}},\ }\bibfield  {title} {\bibinfo {title} {Two-photon
			excitation sets limit to entangled photon pair generation from quantum
			emitters},\ }\href {https://doi.org/10.1103/PhysRevLett.129.193604}
	{\bibfield  {journal} {\bibinfo  {journal} {Phys. Rev. Lett.}\ }\textbf
		{\bibinfo {volume} {129}},\ \bibinfo {pages} {193604} (\bibinfo {year}
		{2022})}\BibitemShut {NoStop}%
	\bibitem [{\citenamefont {Schmolke}\ and\ \citenamefont
		{Lutz}(2022)}]{schmolke22}%
	\BibitemOpen
	\bibfield  {author} {\bibinfo {author} {\bibfnamefont {F.}~\bibnamefont
			{Schmolke}}\ and\ \bibinfo {author} {\bibfnamefont {E.}~\bibnamefont
			{Lutz}},\ }\bibfield  {title} {\bibinfo {title} {Noise-induced quantum
			synchronization},\ }\href {https://doi.org/10.1103/PhysRevLett.129.250601}
	{\bibfield  {journal} {\bibinfo  {journal} {Phys. Rev. Lett.}\ }\textbf
		{\bibinfo {volume} {129}},\ \bibinfo {pages} {250601} (\bibinfo {year}
		{2022})}\BibitemShut {NoStop}%
	\bibitem [{\citenamefont {Vovk}\ and\ \citenamefont {Pichler}(2022)}]{vovk22}%
	\BibitemOpen
	\bibfield  {author} {\bibinfo {author} {\bibfnamefont {T.}~\bibnamefont
			{Vovk}}\ and\ \bibinfo {author} {\bibfnamefont {H.}~\bibnamefont {Pichler}},\
	}\bibfield  {title} {\bibinfo {title} {Entanglement-optimal trajectories of
			many-body quantum markov processes},\ }\href
	{https://doi.org/10.1103/PhysRevLett.128.243601} {\bibfield  {journal}
		{\bibinfo  {journal} {Phys. Rev. Lett.}\ }\textbf {\bibinfo {volume} {128}},\
		\bibinfo {pages} {243601} (\bibinfo {year} {2022})}\BibitemShut {NoStop}%
	\bibitem [{\citenamefont {Vedral}\ \emph {et~al.}(1997)\citenamefont {Vedral},
		\citenamefont {Plenio}, \citenamefont {Rippin},\ and\ \citenamefont
		{Knight}}]{vedral97}%
	\BibitemOpen
	\bibfield  {author} {\bibinfo {author} {\bibfnamefont {V.}~\bibnamefont
			{Vedral}}, \bibinfo {author} {\bibfnamefont {M.~B.}\ \bibnamefont {Plenio}},
		\bibinfo {author} {\bibfnamefont {M.~A.}\ \bibnamefont {Rippin}},\ and\
		\bibinfo {author} {\bibfnamefont {P.~L.}\ \bibnamefont {Knight}},\ }\bibfield
	{title} {\bibinfo {title} {Quantifying entanglement},\ }\href
	{https://doi.org/10.1103/PhysRevLett.78.2275} {\bibfield  {journal} {\bibinfo
			{journal} {Phys. Rev. Lett.}\ }\textbf {\bibinfo {volume} {78}},\ \bibinfo
		{pages} {2275} (\bibinfo {year} {1997})}\BibitemShut {NoStop}%
	\bibitem [{\citenamefont {Rains}(1999)}]{rains99}%
	\BibitemOpen
	\bibfield  {author} {\bibinfo {author} {\bibfnamefont {E.~M.}\ \bibnamefont
			{Rains}},\ }\bibfield  {title} {\bibinfo {title} {Rigorous treatment of
			distillable entanglement},\ }\href {https://doi.org/10.1103/PhysRevA.60.173}
	{\bibfield  {journal} {\bibinfo  {journal} {Phys. Rev. A}\ }\textbf {\bibinfo
			{volume} {60}},\ \bibinfo {pages} {173} (\bibinfo {year} {1999})}\BibitemShut
	{NoStop}%
	\bibitem [{\citenamefont {Vidal}\ and\ \citenamefont {Werner}(2002)}]{vidal02}%
	\BibitemOpen
	\bibfield  {author} {\bibinfo {author} {\bibfnamefont {G.}~\bibnamefont
			{Vidal}}\ and\ \bibinfo {author} {\bibfnamefont {R.~F.}\ \bibnamefont
			{Werner}},\ }\bibfield  {title} {\bibinfo {title} {Computable measure of
			entanglement},\ }\href {https://doi.org/10.1103/PhysRevA.65.032314}
	{\bibfield  {journal} {\bibinfo  {journal} {Phys. Rev. A}\ }\textbf {\bibinfo
			{volume} {65}},\ \bibinfo {pages} {032314} (\bibinfo {year}
		{2002})}\BibitemShut {NoStop}%
	\bibitem [{\citenamefont {Plenio}(2005)}]{plenio05b}%
	\BibitemOpen
	\bibfield  {author} {\bibinfo {author} {\bibfnamefont {M.~B.}\ \bibnamefont
			{Plenio}},\ }\bibfield  {title} {\bibinfo {title} {Logarithmic negativity: A
			full entanglement monotone that is not convex},\ }\href
	{https://doi.org/10.1103/PhysRevLett.95.090503} {\bibfield  {journal}
		{\bibinfo  {journal} {Phys. Rev. Lett.}\ }\textbf {\bibinfo {volume} {95}},\
		\bibinfo {pages} {090503} (\bibinfo {year} {2005})}\BibitemShut {NoStop}%
	\bibitem [{\citenamefont {Peres}(1996)}]{peres96a}%
	\BibitemOpen
	\bibfield  {author} {\bibinfo {author} {\bibfnamefont {A.}~\bibnamefont
			{Peres}},\ }\bibfield  {title} {\bibinfo {title} {Separability criterion for
			density matrices},\ }\href {https://doi.org/10.1103/PhysRevLett.77.1413}
	{\bibfield  {journal} {\bibinfo  {journal} {Phys. Rev. Lett.}\ }\textbf
		{\bibinfo {volume} {77}},\ \bibinfo {pages} {1413} (\bibinfo {year}
		{1996})}\BibitemShut {NoStop}%
	\bibitem [{\citenamefont {Horodecki}\ \emph {et~al.}(1996)\citenamefont
		{Horodecki}, \citenamefont {Horodecki},\ and\ \citenamefont
		{Horodecki}}]{horodecki96}%
	\BibitemOpen
	\bibfield  {author} {\bibinfo {author} {\bibfnamefont {M.}~\bibnamefont
			{Horodecki}}, \bibinfo {author} {\bibfnamefont {P.}~\bibnamefont
			{Horodecki}},\ and\ \bibinfo {author} {\bibfnamefont {R.}~\bibnamefont
			{Horodecki}},\ }\bibfield  {title} {\bibinfo {title} {Separability of mixed
			states: necessary and sufficient conditions},\ }\href
	{https://doi.org/https://doi.org/10.1016/S0375-9601(96)00706-2} {\bibfield
		{journal} {\bibinfo  {journal} {Physics Letters A}\ }\textbf {\bibinfo
			{volume} {223}},\ \bibinfo {pages} {1} (\bibinfo {year} {1996})}\BibitemShut
	{NoStop}%
	\bibitem [{\citenamefont {Horodecki}\ \emph {et~al.}(1998)\citenamefont
		{Horodecki}, \citenamefont {Horodecki},\ and\ \citenamefont
		{Horodecki}}]{horodecki98}%
	\BibitemOpen
	\bibfield  {author} {\bibinfo {author} {\bibfnamefont {M.}~\bibnamefont
			{Horodecki}}, \bibinfo {author} {\bibfnamefont {P.}~\bibnamefont
			{Horodecki}},\ and\ \bibinfo {author} {\bibfnamefont {R.}~\bibnamefont
			{Horodecki}},\ }\bibfield  {title} {\bibinfo {title} {Mixed-state
			entanglement and distillation: Is there a ``bound'' entanglement in
			nature?},\ }\href {https://doi.org/10.1103/PhysRevLett.80.5239} {\bibfield
		{journal} {\bibinfo  {journal} {Phys. Rev. Lett.}\ }\textbf {\bibinfo
			{volume} {80}},\ \bibinfo {pages} {5239} (\bibinfo {year}
		{1998})}\BibitemShut {NoStop}%
	\bibitem [{\citenamefont {Smolin}(2001)}]{smolin01}%
	\BibitemOpen
	\bibfield  {author} {\bibinfo {author} {\bibfnamefont {J.~A.}\ \bibnamefont
			{Smolin}},\ }\bibfield  {title} {\bibinfo {title} {Four-party unlockable
			bound entangled state},\ }\href {https://doi.org/10.1103/PhysRevA.63.032306}
	{\bibfield  {journal} {\bibinfo  {journal} {Phys. Rev. A}\ }\textbf {\bibinfo
			{volume} {63}},\ \bibinfo {pages} {032306} (\bibinfo {year}
		{2001})}\BibitemShut {NoStop}%
	\bibitem [{\citenamefont {DiGuglielmo}\ \emph {et~al.}(2011)\citenamefont
		{DiGuglielmo}, \citenamefont {Samblowski}, \citenamefont {Hage},
		\citenamefont {Pineda}, \citenamefont {Eisert},\ and\ \citenamefont
		{Schnabel}}]{diguglielmo11}%
	\BibitemOpen
	\bibfield  {author} {\bibinfo {author} {\bibfnamefont {J.}~\bibnamefont
			{DiGuglielmo}}, \bibinfo {author} {\bibfnamefont {A.}~\bibnamefont
			{Samblowski}}, \bibinfo {author} {\bibfnamefont {B.}~\bibnamefont {Hage}},
		\bibinfo {author} {\bibfnamefont {C.}~\bibnamefont {Pineda}}, \bibinfo
		{author} {\bibfnamefont {J.}~\bibnamefont {Eisert}},\ and\ \bibinfo {author}
		{\bibfnamefont {R.}~\bibnamefont {Schnabel}},\ }\bibfield  {title} {\bibinfo
		{title} {Experimental unconditional preparation and detection of a continuous
			bound entangled state of light},\ }\href
	{https://doi.org/10.1103/PhysRevLett.107.240503} {\bibfield  {journal}
		{\bibinfo  {journal} {Phys. Rev. Lett.}\ }\textbf {\bibinfo {volume} {107}},\
		\bibinfo {pages} {240503} (\bibinfo {year} {2011})}\BibitemShut {NoStop}%
	\bibitem [{\citenamefont {Hiesmayr}\ and\ \citenamefont
		{L{\"o}ffler}(2013)}]{hiesmayr13}%
	\BibitemOpen
	\bibfield  {author} {\bibinfo {author} {\bibfnamefont {B.~C.}\ \bibnamefont
			{Hiesmayr}}\ and\ \bibinfo {author} {\bibfnamefont {W.}~\bibnamefont
			{L{\"o}ffler}},\ }\bibfield  {title} {\bibinfo {title} {Complementarity
			reveals bound entanglement of two twisted photons},\ }\href@noop {}
	{\bibfield  {journal} {\bibinfo  {journal} {New journal of physics}\ }\textbf
		{\bibinfo {volume} {15}},\ \bibinfo {pages} {083036} (\bibinfo {year}
		{2013})}\BibitemShut {NoStop}%
	\bibitem [{\citenamefont {Sent{\'{i}}s}\ \emph {et~al.}(2018)\citenamefont
		{Sent{\'{i}}s}, \citenamefont {Greiner}, \citenamefont {Shang}, \citenamefont
		{Siewert},\ and\ \citenamefont {Kleinmann}}]{sentis18}%
	\BibitemOpen
	\bibfield  {author} {\bibinfo {author} {\bibfnamefont {G.}~\bibnamefont
			{Sent{\'{i}}s}}, \bibinfo {author} {\bibfnamefont {J.~N.}\ \bibnamefont
			{Greiner}}, \bibinfo {author} {\bibfnamefont {J.}~\bibnamefont {Shang}},
		\bibinfo {author} {\bibfnamefont {J.}~\bibnamefont {Siewert}},\ and\ \bibinfo
		{author} {\bibfnamefont {M.}~\bibnamefont {Kleinmann}},\ }\bibfield  {title}
	{\bibinfo {title} {Bound entangled states fit for robust experimental
			verification},\ }\href {https://doi.org/10.22331/q-2018-12-18-113} {\bibfield
		{journal} {\bibinfo  {journal} {{Quantum}}\ }\textbf {\bibinfo {volume}
			{2}},\ \bibinfo {pages} {113} (\bibinfo {year} {2018})}\BibitemShut {NoStop}%
	\bibitem [{\citenamefont {Gabdulin}\ and\ \citenamefont
		{Mandilara}(2019)}]{gabodulin19}%
	\BibitemOpen
	\bibfield  {author} {\bibinfo {author} {\bibfnamefont {A.}~\bibnamefont
			{Gabdulin}}\ and\ \bibinfo {author} {\bibfnamefont {A.}~\bibnamefont
			{Mandilara}},\ }\bibfield  {title} {\bibinfo {title} {Investigating bound
			entangled two-qutrit states via the best separable approximation},\ }\href
	{https://doi.org/10.1103/PhysRevA.100.062322} {\bibfield  {journal} {\bibinfo
			{journal} {Phys. Rev. A}\ }\textbf {\bibinfo {volume} {100}},\ \bibinfo
		{pages} {062322} (\bibinfo {year} {2019})}\BibitemShut {NoStop}%
	\bibitem [{\citenamefont {Eickbusch}\ \emph {et~al.}(2022)\citenamefont
		{Eickbusch}, \citenamefont {Sivak}, \citenamefont {Ding}, \citenamefont
		{Elder}, \citenamefont {Jha}, \citenamefont {Venkatraman}, \citenamefont
		{Royer}, \citenamefont {Girvin}, \citenamefont {Schoelkopf},\ and\
		\citenamefont {Devoret}}]{eickbusch22}%
	\BibitemOpen
	\bibfield  {author} {\bibinfo {author} {\bibfnamefont {A.}~\bibnamefont
			{Eickbusch}}, \bibinfo {author} {\bibfnamefont {V.}~\bibnamefont {Sivak}},
		\bibinfo {author} {\bibfnamefont {A.~Z.}\ \bibnamefont {Ding}}, \bibinfo
		{author} {\bibfnamefont {S.~S.}\ \bibnamefont {Elder}}, \bibinfo {author}
		{\bibfnamefont {S.~R.}\ \bibnamefont {Jha}}, \bibinfo {author} {\bibfnamefont
			{J.}~\bibnamefont {Venkatraman}}, \bibinfo {author} {\bibfnamefont
			{B.}~\bibnamefont {Royer}}, \bibinfo {author} {\bibfnamefont {S.~M.}\
			\bibnamefont {Girvin}}, \bibinfo {author} {\bibfnamefont {R.~J.}\
			\bibnamefont {Schoelkopf}},\ and\ \bibinfo {author} {\bibfnamefont {M.~H.}\
			\bibnamefont {Devoret}},\ }\bibfield  {title} {\bibinfo {title} {Fast
			universal control of an oscillator with weak dispersive coupling to a
			qubit},\ }\href {https://doi.org/10.1038/s41567-022-01776-9} {\bibfield
		{journal} {\bibinfo  {journal} {Nature Physics}\ }\textbf {\bibinfo {volume}
			{18}},\ \bibinfo {pages} {1464} (\bibinfo {year} {2022})}\BibitemShut
	{NoStop}%
	\bibitem [{\citenamefont {Katz}\ and\ \citenamefont {Monroe}(2023)}]{katz23}%
	\BibitemOpen
	\bibfield  {author} {\bibinfo {author} {\bibfnamefont {O.}~\bibnamefont
			{Katz}}\ and\ \bibinfo {author} {\bibfnamefont {C.}~\bibnamefont {Monroe}},\
	}\bibfield  {title} {\bibinfo {title} {Programmable quantum simulations of
			bosonic systems with trapped ions},\ }\href
	{https://doi.org/10.1103/PhysRevLett.131.033604} {\bibfield  {journal}
		{\bibinfo  {journal} {Phys. Rev. Lett.}\ }\textbf {\bibinfo {volume} {131}},\
		\bibinfo {pages} {033604} (\bibinfo {year} {2023})}\BibitemShut {NoStop}%
	\bibitem [{\citenamefont {Ma}\ \emph {et~al.}(2021)\citenamefont {Ma},
		\citenamefont {Puri}, \citenamefont {Schoelkopf}, \citenamefont {Devoret},
		\citenamefont {Girvin},\ and\ \citenamefont {Jiang}}]{ma21}%
	\BibitemOpen
	\bibfield  {author} {\bibinfo {author} {\bibfnamefont {W.-L.}\ \bibnamefont
			{Ma}}, \bibinfo {author} {\bibfnamefont {S.}~\bibnamefont {Puri}}, \bibinfo
		{author} {\bibfnamefont {R.~J.}\ \bibnamefont {Schoelkopf}}, \bibinfo
		{author} {\bibfnamefont {M.~H.}\ \bibnamefont {Devoret}}, \bibinfo {author}
		{\bibfnamefont {S.}~\bibnamefont {Girvin}},\ and\ \bibinfo {author}
		{\bibfnamefont {L.}~\bibnamefont {Jiang}},\ }\bibfield  {title} {\bibinfo
		{title} {Quantum control of bosonic modes with superconducting circuits},\
	}\href {https://doi.org/https://doi.org/10.1016/j.scib.2021.05.024}
	{\bibfield  {journal} {\bibinfo  {journal} {Science Bulletin}\ }\textbf
		{\bibinfo {volume} {66}},\ \bibinfo {pages} {1789} (\bibinfo {year}
		{2021})}\BibitemShut {NoStop}%
	\bibitem [{\citenamefont {Haroche}\ \emph {et~al.}(2020)\citenamefont
		{Haroche}, \citenamefont {Brune},\ and\ \citenamefont {Raimond}}]{haroche20}%
	\BibitemOpen
	\bibfield  {author} {\bibinfo {author} {\bibfnamefont {S.}~\bibnamefont
			{Haroche}}, \bibinfo {author} {\bibfnamefont {M.}~\bibnamefont {Brune}},\
		and\ \bibinfo {author} {\bibfnamefont {J.}~\bibnamefont {Raimond}},\
	}\bibfield  {title} {\bibinfo {title} {From cavity to circuit quantum
			electrodynamics},\ }\href {https://doi.org/10.1038/s41567-020-0812-1}
	{\bibfield  {journal} {\bibinfo  {journal} {Nature Physics}\ }\textbf
		{\bibinfo {volume} {16}},\ \bibinfo {pages} {243} (\bibinfo {year}
		{2020})}\BibitemShut {NoStop}%
	\bibitem [{\citenamefont {Roszak}\ and\ \citenamefont
		{Cywi\ifmmode~\acute{n}\else \'{n}\fi{}ski}(2015)}]{roszak15}%
	\BibitemOpen
	\bibfield  {author} {\bibinfo {author} {\bibfnamefont {K.}~\bibnamefont
			{Roszak}}\ and\ \bibinfo {author} {\bibfnamefont {L.}~\bibnamefont
			{Cywi\ifmmode~\acute{n}\else \'{n}\fi{}ski}},\ }\bibfield  {title} {\bibinfo
		{title} {Characterization and measurement of qubit-environment-entanglement
			generation during pure dephasing},\ }\href
	{https://doi.org/10.1103/PhysRevA.92.032310} {\bibfield  {journal} {\bibinfo
			{journal} {Phys. Rev. A}\ }\textbf {\bibinfo {volume} {92}},\ \bibinfo
		{pages} {032310} (\bibinfo {year} {2015})}\BibitemShut {NoStop}%
	\bibitem [{\citenamefont {Roszak}(2018)}]{roszak18}%
	\BibitemOpen
	\bibfield  {author} {\bibinfo {author} {\bibfnamefont {K.}~\bibnamefont
			{Roszak}},\ }\bibfield  {title} {\bibinfo {title} {Criteria for
			system-environment entanglement generation for systems of any size in
			pure-dephasing evolutions},\ }\href
	{https://doi.org/10.1103/PhysRevA.98.052344} {\bibfield  {journal} {\bibinfo
			{journal} {Phys. Rev. A}\ }\textbf {\bibinfo {volume} {98}},\ \bibinfo
		{pages} {052344} (\bibinfo {year} {2018})}\BibitemShut {NoStop}%
	\bibitem [{\citenamefont {Roszak}\ \emph {et~al.}(2019)\citenamefont {Roszak},
		\citenamefont {Kwiatkowski},\ and\ \citenamefont {Cywi\ifmmode~\acute{n}\else
			\'{n}\fi{}ski}}]{roszak19a}%
	\BibitemOpen
	\bibfield  {author} {\bibinfo {author} {\bibfnamefont {K.}~\bibnamefont
			{Roszak}}, \bibinfo {author} {\bibfnamefont {D.}~\bibnamefont
			{Kwiatkowski}},\ and\ \bibinfo {author} {\bibfnamefont {L.}~\bibnamefont
			{Cywi\ifmmode~\acute{n}\else \'{n}\fi{}ski}},\ }\bibfield  {title} {\bibinfo
		{title} {How to detect qubit-environment entanglement generated during qubit
			dephasing},\ }\href {https://doi.org/10.1103/PhysRevA.100.022318} {\bibfield
		{journal} {\bibinfo  {journal} {Phys. Rev. A}\ }\textbf {\bibinfo {volume}
			{100}},\ \bibinfo {pages} {022318} (\bibinfo {year} {2019})}\BibitemShut
	{NoStop}%
	\bibitem [{\citenamefont {Rzepkowski}\ and\ \citenamefont
		{Roszak}(2020)}]{rzepkowski20}%
	\BibitemOpen
	\bibfield  {author} {\bibinfo {author} {\bibfnamefont {B.}~\bibnamefont
			{Rzepkowski}}\ and\ \bibinfo {author} {\bibfnamefont {K.}~\bibnamefont
			{Roszak}},\ }\href@noop {} {\bibinfo {title} {A scheme for direct detection
			of qubit-environment entanglement generated during qubit pure dephasing}}
	(\bibinfo {year} {2020}),\ \Eprint {https://arxiv.org/abs/2002.10901}
	{arXiv:2002.10901 [quant-ph]} \BibitemShut {NoStop}%
	\bibitem [{\citenamefont {Strza\l{}ka}\ and\ \citenamefont
		{Roszak}(2021)}]{strzalka21}%
	\BibitemOpen
	\bibfield  {author} {\bibinfo {author} {\bibfnamefont {M.}~\bibnamefont
			{Strza\l{}ka}}\ and\ \bibinfo {author} {\bibfnamefont {K.}~\bibnamefont
			{Roszak}},\ }\bibfield  {title} {\bibinfo {title} {Detection of entanglement
			during pure dephasing evolutions for systems and environments of any size},\
	}\href {https://doi.org/10.1103/PhysRevA.104.042411} {\bibfield  {journal}
		{\bibinfo  {journal} {Phys. Rev. A}\ }\textbf {\bibinfo {volume} {104}},\
		\bibinfo {pages} {042411} (\bibinfo {year} {2021})}\BibitemShut {NoStop}%
	\bibitem [{\citenamefont {Roszak}\ and\ \citenamefont
		{Cywi\ifmmode~\acute{n}\else \'{n}\fi{}ski}(2021)}]{roszak21}%
	\BibitemOpen
	\bibfield  {author} {\bibinfo {author} {\bibfnamefont {K.}~\bibnamefont
			{Roszak}}\ and\ \bibinfo {author} {\bibfnamefont {L.}~\bibnamefont
			{Cywi\ifmmode~\acute{n}\else \'{n}\fi{}ski}},\ }\bibfield  {title} {\bibinfo
		{title} {Qubit-environment-entanglement generation and the spin echo},\
	}\href {https://doi.org/10.1103/PhysRevA.103.032208} {\bibfield  {journal}
		{\bibinfo  {journal} {Phys. Rev. A}\ }\textbf {\bibinfo {volume} {103}},\
		\bibinfo {pages} {032208} (\bibinfo {year} {2021})}\BibitemShut {NoStop}%
	\bibitem [{\citenamefont {Roszak}\ and\ \citenamefont
		{Cywi\ifmmode~\acute{n}\else \'{n}\fi{}ski}(2018)}]{roszak17}%
	\BibitemOpen
	\bibfield  {author} {\bibinfo {author} {\bibfnamefont {K.}~\bibnamefont
			{Roszak}}\ and\ \bibinfo {author} {\bibfnamefont {L.}~\bibnamefont
			{Cywi\ifmmode~\acute{n}\else \'{n}\fi{}ski}},\ }\bibfield  {title} {\bibinfo
		{title} {Equivalence of qubit-environment entanglement and discord generation
			via pure dephasing interactions and the resulting consequences},\ }\href
	{https://doi.org/10.1103/PhysRevA.97.012306} {\bibfield  {journal} {\bibinfo
			{journal} {Phys. Rev. A}\ }\textbf {\bibinfo {volume} {97}},\ \bibinfo
		{pages} {012306} (\bibinfo {year} {2018})}\BibitemShut {NoStop}%
	\bibitem [{\citenamefont {Harlender}\ and\ \citenamefont
		{Roszak}(2022)}]{harlender22}%
	\BibitemOpen
	\bibfield  {author} {\bibinfo {author} {\bibfnamefont {T.}~\bibnamefont
			{Harlender}}\ and\ \bibinfo {author} {\bibfnamefont {K.}~\bibnamefont
			{Roszak}},\ }\bibfield  {title} {\bibinfo {title} {Transfer and teleportation
			of system-environment entanglement},\ }\href
	{https://doi.org/10.1103/PhysRevA.105.012407} {\bibfield  {journal} {\bibinfo
			{journal} {Phys. Rev. A}\ }\textbf {\bibinfo {volume} {105}},\ \bibinfo
		{pages} {012407} (\bibinfo {year} {2022})}\BibitemShut {NoStop}%
	\bibitem [{\citenamefont {Roszak}\ and\ \citenamefont
		{Korbicz}(2023)}]{roszak23}%
	\BibitemOpen
	\bibfield  {author} {\bibinfo {author} {\bibfnamefont {K.}~\bibnamefont
			{Roszak}}\ and\ \bibinfo {author} {\bibfnamefont {J.~K.}\ \bibnamefont
			{Korbicz}},\ }\bibfield  {title} {\bibinfo {title} {Purifying
			teleportation},\ }\href {https://doi.org/10.22331/q-2023-02-16-923}
	{\bibfield  {journal} {\bibinfo  {journal} {{Quantum}}\ }\textbf {\bibinfo
			{volume} {7}},\ \bibinfo {pages} {923} (\bibinfo {year} {2023})}\BibitemShut
	{NoStop}%
	\bibitem [{\citenamefont {Zurek}(2003)}]{zurek03}%
	\BibitemOpen
	\bibfield  {author} {\bibinfo {author} {\bibfnamefont {W.~H.}\ \bibnamefont
			{Zurek}},\ }\bibfield  {title} {\bibinfo {title} {Decoherence, einselection,
			and the quantum origins of the classical},\ }\href
	{https://doi.org/10.1103/RevModPhys.75.715} {\bibfield  {journal} {\bibinfo
			{journal} {Rev. Mod. Phys.}\ }\textbf {\bibinfo {volume} {75}},\ \bibinfo
		{pages} {715} (\bibinfo {year} {2003})}\BibitemShut {NoStop}%
	\bibitem [{\citenamefont {Ollivier}\ \emph {et~al.}(2004)\citenamefont
		{Ollivier}, \citenamefont {Poulin},\ and\ \citenamefont
		{Zurek}}]{ollivier04}%
	\BibitemOpen
	\bibfield  {author} {\bibinfo {author} {\bibfnamefont {H.}~\bibnamefont
			{Ollivier}}, \bibinfo {author} {\bibfnamefont {D.}~\bibnamefont {Poulin}},\
		and\ \bibinfo {author} {\bibfnamefont {W.~H.}\ \bibnamefont {Zurek}},\
	}\bibfield  {title} {\bibinfo {title} {Objective properties from subjective
			quantum states: Environment as a witness},\ }\href
	{https://doi.org/10.1103/PhysRevLett.93.220401} {\bibfield  {journal}
		{\bibinfo  {journal} {Phys. Rev. Lett.}\ }\textbf {\bibinfo {volume} {93}},\
		\bibinfo {pages} {220401} (\bibinfo {year} {2004})}\BibitemShut {NoStop}%
	\bibitem [{\citenamefont {Ollivier}\ \emph {et~al.}(2005)\citenamefont
		{Ollivier}, \citenamefont {Poulin},\ and\ \citenamefont
		{Zurek}}]{ollivier05}%
	\BibitemOpen
	\bibfield  {author} {\bibinfo {author} {\bibfnamefont {H.}~\bibnamefont
			{Ollivier}}, \bibinfo {author} {\bibfnamefont {D.}~\bibnamefont {Poulin}},\
		and\ \bibinfo {author} {\bibfnamefont {W.~H.}\ \bibnamefont {Zurek}},\
	}\bibfield  {title} {\bibinfo {title} {Environment as a witness: Selective
			proliferation of information and emergence of objectivity in a quantum
			universe},\ }\href {https://doi.org/10.1103/PhysRevA.72.042113} {\bibfield
		{journal} {\bibinfo  {journal} {Phys. Rev. A}\ }\textbf {\bibinfo {volume}
			{72}},\ \bibinfo {pages} {042113} (\bibinfo {year} {2005})}\BibitemShut
	{NoStop}%
	\bibitem [{\citenamefont {Zurek}(2009)}]{zurek09}%
	\BibitemOpen
	\bibfield  {author} {\bibinfo {author} {\bibfnamefont {W.~H.}\ \bibnamefont
			{Zurek}},\ }\bibfield  {title} {\bibinfo {title} {Quantum darwinism},\ }\href
	{https://doi.org/10.1038/nphys1202} {\bibfield  {journal} {\bibinfo
			{journal} {Nature physics}\ }\textbf {\bibinfo {volume} {5}},\ \bibinfo
		{pages} {181} (\bibinfo {year} {2009})}\BibitemShut {NoStop}%
	\bibitem [{\citenamefont {Korbicz}\ \emph {et~al.}(2014)\citenamefont
		{Korbicz}, \citenamefont {Horodecki},\ and\ \citenamefont
		{Horodecki}}]{korbicz14}%
	\BibitemOpen
	\bibfield  {author} {\bibinfo {author} {\bibfnamefont {J.~K.}\ \bibnamefont
			{Korbicz}}, \bibinfo {author} {\bibfnamefont {P.}~\bibnamefont {Horodecki}},\
		and\ \bibinfo {author} {\bibfnamefont {R.}~\bibnamefont {Horodecki}},\
	}\bibfield  {title} {\bibinfo {title} {Objectivity in a noisy photonic
			environment through quantum state information broadcasting},\ }\href
	{https://doi.org/10.1103/PhysRevLett.112.120402} {\bibfield  {journal}
		{\bibinfo  {journal} {Phys. Rev. Lett.}\ }\textbf {\bibinfo {volume} {112}},\
		\bibinfo {pages} {120402} (\bibinfo {year} {2014})}\BibitemShut {NoStop}%
	\bibitem [{\citenamefont {Horodecki}\ \emph {et~al.}(2015)\citenamefont
		{Horodecki}, \citenamefont {Korbicz},\ and\ \citenamefont
		{Horodecki}}]{horodecki15}%
	\BibitemOpen
	\bibfield  {author} {\bibinfo {author} {\bibfnamefont {R.}~\bibnamefont
			{Horodecki}}, \bibinfo {author} {\bibfnamefont {J.~K.}\ \bibnamefont
			{Korbicz}},\ and\ \bibinfo {author} {\bibfnamefont {P.}~\bibnamefont
			{Horodecki}},\ }\bibfield  {title} {\bibinfo {title} {Quantum origins of
			objectivity},\ }\href {https://doi.org/10.1103/PhysRevA.91.032122} {\bibfield
		{journal} {\bibinfo  {journal} {Phys. Rev. A}\ }\textbf {\bibinfo {volume}
			{91}},\ \bibinfo {pages} {032122} (\bibinfo {year} {2015})}\BibitemShut
	{NoStop}%
	\bibitem [{\citenamefont {Mironowicz}\ \emph {et~al.}(2017)\citenamefont
		{Mironowicz}, \citenamefont {Korbicz},\ and\ \citenamefont
		{Horodecki}}]{mironowicz17}%
	\BibitemOpen
	\bibfield  {author} {\bibinfo {author} {\bibfnamefont {P.}~\bibnamefont
			{Mironowicz}}, \bibinfo {author} {\bibfnamefont {J.~K.}\ \bibnamefont
			{Korbicz}},\ and\ \bibinfo {author} {\bibfnamefont {P.}~\bibnamefont
			{Horodecki}},\ }\bibfield  {title} {\bibinfo {title} {Monitoring of the
			process of system information broadcasting in time},\ }\href
	{https://doi.org/10.1103/PhysRevLett.118.150501} {\bibfield  {journal}
		{\bibinfo  {journal} {Phys. Rev. Lett.}\ }\textbf {\bibinfo {volume} {118}},\
		\bibinfo {pages} {150501} (\bibinfo {year} {2017})}\BibitemShut {NoStop}%
	\bibitem [{\citenamefont {Lampo}\ \emph {et~al.}(2017)\citenamefont {Lampo},
		\citenamefont {Tuziemski}, \citenamefont {Lewenstein},\ and\ \citenamefont
		{Korbicz}}]{lampo17}%
	\BibitemOpen
	\bibfield  {author} {\bibinfo {author} {\bibfnamefont {A.}~\bibnamefont
			{Lampo}}, \bibinfo {author} {\bibfnamefont {J.}~\bibnamefont {Tuziemski}},
		\bibinfo {author} {\bibfnamefont {M.}~\bibnamefont {Lewenstein}},\ and\
		\bibinfo {author} {\bibfnamefont {J.~K.}\ \bibnamefont {Korbicz}},\
	}\bibfield  {title} {\bibinfo {title} {Objectivity in the non-markovian
			spin-boson model},\ }\href {https://doi.org/10.1103/PhysRevA.96.012120}
	{\bibfield  {journal} {\bibinfo  {journal} {Phys. Rev. A}\ }\textbf {\bibinfo
			{volume} {96}},\ \bibinfo {pages} {012120} (\bibinfo {year}
		{2017})}\BibitemShut {NoStop}%
	\bibitem [{\citenamefont {Roszak}\ and\ \citenamefont
		{Korbicz}(2019)}]{roszak19b}%
	\BibitemOpen
	\bibfield  {author} {\bibinfo {author} {\bibfnamefont {K.}~\bibnamefont
			{Roszak}}\ and\ \bibinfo {author} {\bibfnamefont {J.~K.}\ \bibnamefont
			{Korbicz}},\ }\bibfield  {title} {\bibinfo {title} {Entanglement and
			objectivity in pure dephasing models},\ }\href
	{https://doi.org/10.1103/PhysRevA.100.062127} {\bibfield  {journal} {\bibinfo
			{journal} {Phys. Rev. A}\ }\textbf {\bibinfo {volume} {100}},\ \bibinfo
		{pages} {062127} (\bibinfo {year} {2019})}\BibitemShut {NoStop}%
	\bibitem [{\citenamefont {Lorenzo}\ \emph {et~al.}(2020)\citenamefont
		{Lorenzo}, \citenamefont {Paternostro},\ and\ \citenamefont
		{Palma}}]{lorenzo20}%
	\BibitemOpen
	\bibfield  {author} {\bibinfo {author} {\bibfnamefont {S.}~\bibnamefont
			{Lorenzo}}, \bibinfo {author} {\bibfnamefont {M.}~\bibnamefont
			{Paternostro}},\ and\ \bibinfo {author} {\bibfnamefont {G.~M.}\ \bibnamefont
			{Palma}},\ }\bibfield  {title} {\bibinfo {title} {Anti-zeno-based dynamical
			control of the unfolding of quantum darwinism},\ }\href
	{https://doi.org/10.1103/PhysRevResearch.2.013164} {\bibfield  {journal}
		{\bibinfo  {journal} {Phys. Rev. Res.}\ }\textbf {\bibinfo {volume} {2}},\
		\bibinfo {pages} {013164} (\bibinfo {year} {2020})}\BibitemShut {NoStop}%
	\bibitem [{\citenamefont {Roszak}\ and\ \citenamefont
		{Korbicz}(2020)}]{roszak20a}%
	\BibitemOpen
	\bibfield  {author} {\bibinfo {author} {\bibfnamefont {K.}~\bibnamefont
			{Roszak}}\ and\ \bibinfo {author} {\bibfnamefont {J.~K.}\ \bibnamefont
			{Korbicz}},\ }\bibfield  {title} {\bibinfo {title} {Glimpse of objectivity in
			bipartite systems for nonentangling pure dephasing evolutions},\ }\href
	{https://doi.org/10.1103/PhysRevA.101.052120} {\bibfield  {journal} {\bibinfo
			{journal} {Phys. Rev. A}\ }\textbf {\bibinfo {volume} {101}},\ \bibinfo
		{pages} {052120} (\bibinfo {year} {2020})}\BibitemShut {NoStop}%
	\bibitem [{\citenamefont {Baldij\~ao}\ \emph {et~al.}(2021)\citenamefont
		{Baldij\~ao}, \citenamefont {Wagner}, \citenamefont {Duarte}, \citenamefont
		{Amaral},\ and\ \citenamefont {Cunha}}]{baldijao21}%
	\BibitemOpen
	\bibfield  {author} {\bibinfo {author} {\bibfnamefont {R.~D.}\ \bibnamefont
			{Baldij\~ao}}, \bibinfo {author} {\bibfnamefont {R.}~\bibnamefont {Wagner}},
		\bibinfo {author} {\bibfnamefont {C.}~\bibnamefont {Duarte}}, \bibinfo
		{author} {\bibfnamefont {B.}~\bibnamefont {Amaral}},\ and\ \bibinfo {author}
		{\bibfnamefont {M.~T.}\ \bibnamefont {Cunha}},\ }\bibfield  {title} {\bibinfo
		{title} {Emergence of noncontextuality under quantum darwinism},\ }\href
	{https://doi.org/10.1103/PRXQuantum.2.030351} {\bibfield  {journal} {\bibinfo
			{journal} {PRX Quantum}\ }\textbf {\bibinfo {volume} {2}},\ \bibinfo {pages}
		{030351} (\bibinfo {year} {2021})}\BibitemShut {NoStop}%
	\bibitem [{\citenamefont {Borri}\ \emph {et~al.}(2001)\citenamefont {Borri},
		\citenamefont {Langbein}, \citenamefont {Schneider}, \citenamefont {Woggon},
		\citenamefont {Sellin}, \citenamefont {Ouyang},\ and\ \citenamefont
		{Bimberg}}]{borri01}%
	\BibitemOpen
	\bibfield  {author} {\bibinfo {author} {\bibfnamefont {P.}~\bibnamefont
			{Borri}}, \bibinfo {author} {\bibfnamefont {W.}~\bibnamefont {Langbein}},
		\bibinfo {author} {\bibfnamefont {S.}~\bibnamefont {Schneider}}, \bibinfo
		{author} {\bibfnamefont {U.}~\bibnamefont {Woggon}}, \bibinfo {author}
		{\bibfnamefont {R.~L.}\ \bibnamefont {Sellin}}, \bibinfo {author}
		{\bibfnamefont {D.}~\bibnamefont {Ouyang}},\ and\ \bibinfo {author}
		{\bibfnamefont {D.}~\bibnamefont {Bimberg}},\ }\bibfield  {title} {\bibinfo
		{title} {Ultralong dephasing time in {InGaAs} quantum dots},\ }\href@noop {}
	{\ \textbf {\bibinfo {volume} {87}},\ \bibinfo {pages} {157401} (\bibinfo
		{year} {2001})}\BibitemShut {NoStop}%
	\bibitem [{\citenamefont {Vagov}\ \emph {et~al.}(2003)\citenamefont {Vagov},
		\citenamefont {Axt},\ and\ \citenamefont {Kuhn}}]{vagov03}%
	\BibitemOpen
	\bibfield  {author} {\bibinfo {author} {\bibfnamefont {A.}~\bibnamefont
			{Vagov}}, \bibinfo {author} {\bibfnamefont {V.~M.}\ \bibnamefont {Axt}},\
		and\ \bibinfo {author} {\bibfnamefont {T.}~\bibnamefont {Kuhn}},\ }\bibfield
	{title} {\bibinfo {title} {Impact of pure dephasing on the nonlinear optical
			response of single quantum dots and dot ensembles},\ }\href@noop {} {\
		\textbf {\bibinfo {volume} {67}},\ \bibinfo {pages} {115338} (\bibinfo {year}
		{2003})}\BibitemShut {NoStop}%
	\bibitem [{\citenamefont {Vagov}\ \emph {et~al.}(2004)\citenamefont {Vagov},
		\citenamefont {Axt}, \citenamefont {Kuhn}, \citenamefont {Langbein},
		\citenamefont {Borri},\ and\ \citenamefont {Woggon}}]{vagov04}%
	\BibitemOpen
	\bibfield  {author} {\bibinfo {author} {\bibfnamefont {A.}~\bibnamefont
			{Vagov}}, \bibinfo {author} {\bibfnamefont {V.~M.}\ \bibnamefont {Axt}},
		\bibinfo {author} {\bibfnamefont {T.}~\bibnamefont {Kuhn}}, \bibinfo {author}
		{\bibfnamefont {W.}~\bibnamefont {Langbein}}, \bibinfo {author}
		{\bibfnamefont {P.}~\bibnamefont {Borri}},\ and\ \bibinfo {author}
		{\bibfnamefont {U.}~\bibnamefont {Woggon}},\ }\bibfield  {title} {\bibinfo
		{title} {Nonmonotonous temperature dependence of the initial decoherence in
			quantum dots},\ }\href@noop {} {\ \textbf {\bibinfo {volume} {70}},\ \bibinfo
		{pages} {201305(R)} (\bibinfo {year} {2004})}\BibitemShut {NoStop}%
	\bibitem [{\citenamefont {Gl\"assl}\ \emph {et~al.}(2013)\citenamefont
		{Gl\"assl}, \citenamefont {Barth},\ and\ \citenamefont {Axt}}]{glassl13}%
	\BibitemOpen
	\bibfield  {author} {\bibinfo {author} {\bibfnamefont {M.}~\bibnamefont
			{Gl\"assl}}, \bibinfo {author} {\bibfnamefont {A.~M.}\ \bibnamefont
			{Barth}},\ and\ \bibinfo {author} {\bibfnamefont {V.~M.}\ \bibnamefont
			{Axt}},\ }\bibfield  {title} {\bibinfo {title} {Proposed robust and
			high-fidelity preparation of excitons and biexcitons in semiconductor quantum
			dots making active use of phonons},\ }\href
	{https://doi.org/10.1103/PhysRevLett.110.147401} {\bibfield  {journal}
		{\bibinfo  {journal} {Phys. Rev. Lett.}\ }\textbf {\bibinfo {volume} {110}},\
		\bibinfo {pages} {147401} (\bibinfo {year} {2013})}\BibitemShut {NoStop}%
	\bibitem [{\citenamefont {Tahara}\ \emph {et~al.}(2014)\citenamefont {Tahara},
		\citenamefont {Ogawa}, \citenamefont {Minami}, \citenamefont {Akahane},\ and\
		\citenamefont {Sasaki}}]{tahara14}%
	\BibitemOpen
	\bibfield  {author} {\bibinfo {author} {\bibfnamefont {H.}~\bibnamefont
			{Tahara}}, \bibinfo {author} {\bibfnamefont {Y.}~\bibnamefont {Ogawa}},
		\bibinfo {author} {\bibfnamefont {F.}~\bibnamefont {Minami}}, \bibinfo
		{author} {\bibfnamefont {K.}~\bibnamefont {Akahane}},\ and\ \bibinfo {author}
		{\bibfnamefont {M.}~\bibnamefont {Sasaki}},\ }\bibfield  {title} {\bibinfo
		{title} {Long-time correlation in non-markovian dephasing of an
			exciton-phonon system in inas quantum dots},\ }\href
	{https://doi.org/10.1103/PhysRevLett.112.147404} {\bibfield  {journal}
		{\bibinfo  {journal} {Phys. Rev. Lett.}\ }\textbf {\bibinfo {volume} {112}},\
		\bibinfo {pages} {147404} (\bibinfo {year} {2014})}\BibitemShut {NoStop}%
	\bibitem [{\citenamefont {Salamon}\ and\ \citenamefont
		{Roszak}(2017)}]{salamon17}%
	\BibitemOpen
	\bibfield  {author} {\bibinfo {author} {\bibfnamefont {T.}~\bibnamefont
			{Salamon}}\ and\ \bibinfo {author} {\bibfnamefont {K.}~\bibnamefont
			{Roszak}},\ }\bibfield  {title} {\bibinfo {title} {Entanglement generation
			between a charge qubit and its bosonic environment during pure dephasing:
			Dependence on the environment size},\ }\href
	{https://doi.org/10.1103/PhysRevA.96.032333} {\bibfield  {journal} {\bibinfo
			{journal} {Phys. Rev. A}\ }\textbf {\bibinfo {volume} {96}},\ \bibinfo
		{pages} {032333} (\bibinfo {year} {2017})}\BibitemShut {NoStop}%
	\bibitem [{\citenamefont {Seidelmann}\ \emph {et~al.}(2019)\citenamefont
		{Seidelmann}, \citenamefont {Ungar}, \citenamefont {Barth}, \citenamefont
		{Vagov}, \citenamefont {Axt}, \citenamefont {Cygorek},\ and\ \citenamefont
		{Kuhn}}]{seidelmann19}%
	\BibitemOpen
	\bibfield  {author} {\bibinfo {author} {\bibfnamefont {T.}~\bibnamefont
			{Seidelmann}}, \bibinfo {author} {\bibfnamefont {F.}~\bibnamefont {Ungar}},
		\bibinfo {author} {\bibfnamefont {A.~M.}\ \bibnamefont {Barth}}, \bibinfo
		{author} {\bibfnamefont {A.}~\bibnamefont {Vagov}}, \bibinfo {author}
		{\bibfnamefont {V.~M.}\ \bibnamefont {Axt}}, \bibinfo {author} {\bibfnamefont
			{M.}~\bibnamefont {Cygorek}},\ and\ \bibinfo {author} {\bibfnamefont
			{T.}~\bibnamefont {Kuhn}},\ }\bibfield  {title} {\bibinfo {title}
		{Phonon-induced enhancement of photon entanglement in quantum dot-cavity
			systems},\ }\href {https://doi.org/10.1103/PhysRevLett.123.137401} {\bibfield
		{journal} {\bibinfo  {journal} {Phys. Rev. Lett.}\ }\textbf {\bibinfo
			{volume} {123}},\ \bibinfo {pages} {137401} (\bibinfo {year}
		{2019})}\BibitemShut {NoStop}%
	\bibitem [{\citenamefont {Zhao}\ \emph {et~al.}(2012)\citenamefont {Zhao},
		\citenamefont {Ho},\ and\ \citenamefont {Liu}}]{zhao12}%
	\BibitemOpen
	\bibfield  {author} {\bibinfo {author} {\bibfnamefont {N.}~\bibnamefont
			{Zhao}}, \bibinfo {author} {\bibfnamefont {S.-W.}\ \bibnamefont {Ho}},\ and\
		\bibinfo {author} {\bibfnamefont {R.-B.}\ \bibnamefont {Liu}},\ }\bibfield
	{title} {\bibinfo {title} {Decoherence and dynamical decoupling control of
			nitrogen vacancy center electron spins in nuclear spin baths},\ }\href
	{https://doi.org/10.1103/PhysRevB.85.115303} {\bibfield  {journal} {\bibinfo
			{journal} {Phys. Rev. B}\ }\textbf {\bibinfo {volume} {85}},\ \bibinfo
		{pages} {115303} (\bibinfo {year} {2012})}\BibitemShut {NoStop}%
	\bibitem [{\citenamefont {Kwiatkowski}\ and\ \citenamefont
		{Cywi\ifmmode~\acute{n}\else \'{n}\fi{}ski}(2018)}]{kwiatkowski18}%
	\BibitemOpen
	\bibfield  {author} {\bibinfo {author} {\bibfnamefont {D.}~\bibnamefont
			{Kwiatkowski}}\ and\ \bibinfo {author} {\bibfnamefont {L.}~\bibnamefont
			{Cywi\ifmmode~\acute{n}\else \'{n}\fi{}ski}},\ }\bibfield  {title} {\bibinfo
		{title} {Decoherence of two entangled spin qubits coupled to an interacting
			sparse nuclear spin bath: Application to nitrogen vacancy centers},\ }\href
	{https://doi.org/10.1103/PhysRevB.98.155202} {\bibfield  {journal} {\bibinfo
			{journal} {Phys. Rev. B}\ }\textbf {\bibinfo {volume} {98}},\ \bibinfo
		{pages} {155202} (\bibinfo {year} {2018})}\BibitemShut {NoStop}%
	\bibitem [{\citenamefont {Bartling}\ \emph {et~al.}(2022)\citenamefont
		{Bartling}, \citenamefont {Abobeih}, \citenamefont {Pingault}, \citenamefont
		{Degen}, \citenamefont {Loenen}, \citenamefont {Bradley}, \citenamefont
		{Randall}, \citenamefont {Markham}, \citenamefont {Twitchen},\ and\
		\citenamefont {Taminiau}}]{bartling22}%
	\BibitemOpen
	\bibfield  {author} {\bibinfo {author} {\bibfnamefont {H.~P.}\ \bibnamefont
			{Bartling}}, \bibinfo {author} {\bibfnamefont {M.~H.}\ \bibnamefont
			{Abobeih}}, \bibinfo {author} {\bibfnamefont {B.}~\bibnamefont {Pingault}},
		\bibinfo {author} {\bibfnamefont {M.~J.}\ \bibnamefont {Degen}}, \bibinfo
		{author} {\bibfnamefont {S.~J.~H.}\ \bibnamefont {Loenen}}, \bibinfo {author}
		{\bibfnamefont {C.~E.}\ \bibnamefont {Bradley}}, \bibinfo {author}
		{\bibfnamefont {J.}~\bibnamefont {Randall}}, \bibinfo {author} {\bibfnamefont
			{M.}~\bibnamefont {Markham}}, \bibinfo {author} {\bibfnamefont {D.~J.}\
			\bibnamefont {Twitchen}},\ and\ \bibinfo {author} {\bibfnamefont {T.~H.}\
			\bibnamefont {Taminiau}},\ }\bibfield  {title} {\bibinfo {title}
		{Entanglement of spin-pair qubits with intrinsic dephasing times exceeding a
			minute},\ }\href {https://doi.org/10.1103/PhysRevX.12.011048} {\bibfield
		{journal} {\bibinfo  {journal} {Phys. Rev. X}\ }\textbf {\bibinfo {volume}
			{12}},\ \bibinfo {pages} {011048} (\bibinfo {year} {2022})}\BibitemShut
	{NoStop}%
	\bibitem [{\citenamefont {Bayliss}\ \emph {et~al.}(2022)\citenamefont
		{Bayliss}, \citenamefont {Deb}, \citenamefont {Laorenza}, \citenamefont
		{Onizhuk}, \citenamefont {Galli}, \citenamefont {Freedman},\ and\
		\citenamefont {Awschalom}}]{bayliss22}%
	\BibitemOpen
	\bibfield  {author} {\bibinfo {author} {\bibfnamefont {S.~L.}\ \bibnamefont
			{Bayliss}}, \bibinfo {author} {\bibfnamefont {P.}~\bibnamefont {Deb}},
		\bibinfo {author} {\bibfnamefont {D.~W.}\ \bibnamefont {Laorenza}}, \bibinfo
		{author} {\bibfnamefont {M.}~\bibnamefont {Onizhuk}}, \bibinfo {author}
		{\bibfnamefont {G.}~\bibnamefont {Galli}}, \bibinfo {author} {\bibfnamefont
			{D.~E.}\ \bibnamefont {Freedman}},\ and\ \bibinfo {author} {\bibfnamefont
			{D.~D.}\ \bibnamefont {Awschalom}},\ }\bibfield  {title} {\bibinfo {title}
		{Enhancing spin coherence in optically addressable molecular qubits through
			host-matrix control},\ }\href {https://doi.org/10.1103/PhysRevX.12.031028}
	{\bibfield  {journal} {\bibinfo  {journal} {Phys. Rev. X}\ }\textbf {\bibinfo
			{volume} {12}},\ \bibinfo {pages} {031028} (\bibinfo {year}
		{2022})}\BibitemShut {NoStop}%
	\bibitem [{\citenamefont {Wang}\ \emph {et~al.}(2022)\citenamefont {Wang},
		\citenamefont {Liu}, \citenamefont {Fan}, \citenamefont {Feng}, \citenamefont
		{Leong}, \citenamefont {Finkler}, \citenamefont {Denisenko}, \citenamefont
		{Wrachtrup}, \citenamefont {Li},\ and\ \citenamefont {Liu}}]{wang22}%
	\BibitemOpen
	\bibfield  {author} {\bibinfo {author} {\bibfnamefont {N.}~\bibnamefont
			{Wang}}, \bibinfo {author} {\bibfnamefont {C.-F.}\ \bibnamefont {Liu}},
		\bibinfo {author} {\bibfnamefont {J.-W.}\ \bibnamefont {Fan}}, \bibinfo
		{author} {\bibfnamefont {X.}~\bibnamefont {Feng}}, \bibinfo {author}
		{\bibfnamefont {W.-H.}\ \bibnamefont {Leong}}, \bibinfo {author}
		{\bibfnamefont {A.}~\bibnamefont {Finkler}}, \bibinfo {author} {\bibfnamefont
			{A.}~\bibnamefont {Denisenko}}, \bibinfo {author} {\bibfnamefont
			{J.}~\bibnamefont {Wrachtrup}}, \bibinfo {author} {\bibfnamefont
			{Q.}~\bibnamefont {Li}},\ and\ \bibinfo {author} {\bibfnamefont {R.-B.}\
			\bibnamefont {Liu}},\ }\bibfield  {title} {\bibinfo {title} {Zero-field
			magnetometry using hyperfine-biased nitrogen-vacancy centers near diamond
			surfaces},\ }\href {https://doi.org/10.1103/PhysRevResearch.4.013098}
	{\bibfield  {journal} {\bibinfo  {journal} {Phys. Rev. Res.}\ }\textbf
		{\bibinfo {volume} {4}},\ \bibinfo {pages} {013098} (\bibinfo {year}
		{2022})}\BibitemShut {NoStop}%
	\bibitem [{\citenamefont {Onizhuk}\ and\ \citenamefont
		{Galli}(2023)}]{onizhuk23}%
	\BibitemOpen
	\bibfield  {author} {\bibinfo {author} {\bibfnamefont {M.}~\bibnamefont
			{Onizhuk}}\ and\ \bibinfo {author} {\bibfnamefont {G.}~\bibnamefont
			{Galli}},\ }\bibfield  {title} {\bibinfo {title} {Bath-limited dynamics of
			nuclear spins in solid-state spin platforms},\ }\href
	{https://doi.org/10.1103/PhysRevB.108.075306} {\bibfield  {journal} {\bibinfo
			{journal} {Phys. Rev. B}\ }\textbf {\bibinfo {volume} {108}},\ \bibinfo
		{pages} {075306} (\bibinfo {year} {2023})}\BibitemShut {NoStop}%
	\bibitem [{\citenamefont {Roszak}(2020)}]{roszak20}%
	\BibitemOpen
	\bibfield  {author} {\bibinfo {author} {\bibfnamefont {K.}~\bibnamefont
			{Roszak}},\ }\bibfield  {title} {\bibinfo {title} {Measure of
			qubit-environment entanglement for pure dephasing evolutions},\ }\href
	{https://doi.org/10.1103/PhysRevResearch.2.043062} {\bibfield  {journal}
		{\bibinfo  {journal} {Phys. Rev. Research}\ }\textbf {\bibinfo {volume}
			{2}},\ \bibinfo {pages} {043062} (\bibinfo {year} {2020})}\BibitemShut
	{NoStop}%
	\bibitem [{\citenamefont {Touzard}\ \emph {et~al.}(2019)\citenamefont
		{Touzard}, \citenamefont {Kou}, \citenamefont {Frattini}, \citenamefont
		{Sivak}, \citenamefont {Puri}, \citenamefont {Grimm}, \citenamefont
		{Frunzio}, \citenamefont {Shankar},\ and\ \citenamefont
		{Devoret}}]{touzard19}%
	\BibitemOpen
	\bibfield  {author} {\bibinfo {author} {\bibfnamefont {S.}~\bibnamefont
			{Touzard}}, \bibinfo {author} {\bibfnamefont {A.}~\bibnamefont {Kou}},
		\bibinfo {author} {\bibfnamefont {N.~E.}\ \bibnamefont {Frattini}}, \bibinfo
		{author} {\bibfnamefont {V.~V.}\ \bibnamefont {Sivak}}, \bibinfo {author}
		{\bibfnamefont {S.}~\bibnamefont {Puri}}, \bibinfo {author} {\bibfnamefont
			{A.}~\bibnamefont {Grimm}}, \bibinfo {author} {\bibfnamefont
			{L.}~\bibnamefont {Frunzio}}, \bibinfo {author} {\bibfnamefont
			{S.}~\bibnamefont {Shankar}},\ and\ \bibinfo {author} {\bibfnamefont {M.~H.}\
			\bibnamefont {Devoret}},\ }\bibfield  {title} {\bibinfo {title} {Gated
			conditional displacement readout of superconducting qubits},\ }\href
	{https://doi.org/10.1103/PhysRevLett.122.080502} {\bibfield  {journal}
		{\bibinfo  {journal} {Phys. Rev. Lett.}\ }\textbf {\bibinfo {volume} {122}},\
		\bibinfo {pages} {080502} (\bibinfo {year} {2019})}\BibitemShut {NoStop}%
	\bibitem [{\citenamefont {Campagne-Ibarcq}\ \emph {et~al.}(2020)\citenamefont
		{Campagne-Ibarcq}, \citenamefont {Eickbusch}, \citenamefont {Touzard},
		\citenamefont {Zalys-Geller}, \citenamefont {Frattini}, \citenamefont
		{Sivak}, \citenamefont {Reinhold}, \citenamefont {Puri}, \citenamefont
		{Shankar}, \citenamefont {Schoelkopf} \emph {et~al.}}]{campagne20}%
	\BibitemOpen
	\bibfield  {author} {\bibinfo {author} {\bibfnamefont {P.}~\bibnamefont
			{Campagne-Ibarcq}}, \bibinfo {author} {\bibfnamefont {A.}~\bibnamefont
			{Eickbusch}}, \bibinfo {author} {\bibfnamefont {S.}~\bibnamefont {Touzard}},
		\bibinfo {author} {\bibfnamefont {E.}~\bibnamefont {Zalys-Geller}}, \bibinfo
		{author} {\bibfnamefont {N.~E.}\ \bibnamefont {Frattini}}, \bibinfo {author}
		{\bibfnamefont {V.~V.}\ \bibnamefont {Sivak}}, \bibinfo {author}
		{\bibfnamefont {P.}~\bibnamefont {Reinhold}}, \bibinfo {author}
		{\bibfnamefont {S.}~\bibnamefont {Puri}}, \bibinfo {author} {\bibfnamefont
			{S.}~\bibnamefont {Shankar}}, \bibinfo {author} {\bibfnamefont {R.~J.}\
			\bibnamefont {Schoelkopf}}, \emph {et~al.},\ }\bibfield  {title} {\bibinfo
		{title} {Quantum error correction of a qubit encoded in grid states of an
			oscillator},\ }\href
	{https://doi.org/https://doi.org/10.1038/s41586-020-2603-3} {\bibfield
		{journal} {\bibinfo  {journal} {Nature}\ }\textbf {\bibinfo {volume} {584}},\
		\bibinfo {pages} {368} (\bibinfo {year} {2020})}\BibitemShut {NoStop}%
	\bibitem [{\citenamefont {Gao}\ \emph {et~al.}(2021)\citenamefont {Gao},
		\citenamefont {Rol}, \citenamefont {Touzard},\ and\ \citenamefont
		{Wang}}]{gao21}%
	\BibitemOpen
	\bibfield  {author} {\bibinfo {author} {\bibfnamefont {Y.~Y.}\ \bibnamefont
			{Gao}}, \bibinfo {author} {\bibfnamefont {M.~A.}\ \bibnamefont {Rol}},
		\bibinfo {author} {\bibfnamefont {S.}~\bibnamefont {Touzard}},\ and\ \bibinfo
		{author} {\bibfnamefont {C.}~\bibnamefont {Wang}},\ }\bibfield  {title}
	{\bibinfo {title} {Practical guide for building superconducting quantum
			devices},\ }\href {https://doi.org/10.1103/PRXQuantum.2.040202} {\bibfield
		{journal} {\bibinfo  {journal} {PRX Quantum}\ }\textbf {\bibinfo {volume}
			{2}},\ \bibinfo {pages} {040202} (\bibinfo {year} {2021})}\BibitemShut
	{NoStop}%
	\bibitem [{\citenamefont {Delaney}\ \emph {et~al.}(2022)\citenamefont
		{Delaney}, \citenamefont {Urmey}, \citenamefont {Mittal}, \citenamefont
		{Brubaker}, \citenamefont {Kindem}, \citenamefont {Burns}, \citenamefont
		{Regal},\ and\ \citenamefont {Lehnert}}]{delaney22}%
	\BibitemOpen
	\bibfield  {author} {\bibinfo {author} {\bibfnamefont {R.}~\bibnamefont
			{Delaney}}, \bibinfo {author} {\bibfnamefont {M.}~\bibnamefont {Urmey}},
		\bibinfo {author} {\bibfnamefont {S.}~\bibnamefont {Mittal}}, \bibinfo
		{author} {\bibfnamefont {B.}~\bibnamefont {Brubaker}}, \bibinfo {author}
		{\bibfnamefont {J.}~\bibnamefont {Kindem}}, \bibinfo {author} {\bibfnamefont
			{P.}~\bibnamefont {Burns}}, \bibinfo {author} {\bibfnamefont
			{C.}~\bibnamefont {Regal}},\ and\ \bibinfo {author} {\bibfnamefont
			{K.}~\bibnamefont {Lehnert}},\ }\bibfield  {title} {\bibinfo {title}
		{Superconducting-qubit readout via low-backaction electro-optic
			transduction},\ }\href {https://doi.org/10.1038/s41586-022-04720-2}
	{\bibfield  {journal} {\bibinfo  {journal} {Nature}\ }\textbf {\bibinfo
			{volume} {606}},\ \bibinfo {pages} {489} (\bibinfo {year}
		{2022})}\BibitemShut {NoStop}%
	\bibitem [{\citenamefont {Hassani}\ \emph {et~al.}(2023)\citenamefont
		{Hassani}, \citenamefont {Peruzzo}, \citenamefont {Kapoor}, \citenamefont
		{Trioni}, \citenamefont {Zemlicka},\ and\ \citenamefont {Fink}}]{hassani23}%
	\BibitemOpen
	\bibfield  {author} {\bibinfo {author} {\bibfnamefont {F.}~\bibnamefont
			{Hassani}}, \bibinfo {author} {\bibfnamefont {M.}~\bibnamefont {Peruzzo}},
		\bibinfo {author} {\bibfnamefont {L.}~\bibnamefont {Kapoor}}, \bibinfo
		{author} {\bibfnamefont {A.}~\bibnamefont {Trioni}}, \bibinfo {author}
		{\bibfnamefont {M.}~\bibnamefont {Zemlicka}},\ and\ \bibinfo {author}
		{\bibfnamefont {J.~M.}\ \bibnamefont {Fink}},\ }\bibfield  {title} {\bibinfo
		{title} {Inductively shunted transmons exhibit noise insensitive plasmon
			states and a fluxon decay exceeding 3 hours},\ }\href
	{https://doi.org/10.1038/s41467-023-39656-2} {\bibfield  {journal} {\bibinfo
			{journal} {Nature Communications}\ }\textbf {\bibinfo {volume} {14}},\
		\bibinfo {pages} {3968} (\bibinfo {year} {2023})}\BibitemShut {NoStop}%
	\bibitem [{\citenamefont {Leibfried}\ \emph {et~al.}(2003)\citenamefont
		{Leibfried}, \citenamefont {Blatt}, \citenamefont {Monroe},\ and\
		\citenamefont {Wineland}}]{leibfried03}%
	\BibitemOpen
	\bibfield  {author} {\bibinfo {author} {\bibfnamefont {D.}~\bibnamefont
			{Leibfried}}, \bibinfo {author} {\bibfnamefont {R.}~\bibnamefont {Blatt}},
		\bibinfo {author} {\bibfnamefont {C.}~\bibnamefont {Monroe}},\ and\ \bibinfo
		{author} {\bibfnamefont {D.}~\bibnamefont {Wineland}},\ }\bibfield  {title}
	{\bibinfo {title} {Quantum dynamics of single trapped ions},\ }\href
	{https://doi.org/10.1103/RevModPhys.75.281} {\bibfield  {journal} {\bibinfo
			{journal} {Rev. Mod. Phys.}\ }\textbf {\bibinfo {volume} {75}},\ \bibinfo
		{pages} {281} (\bibinfo {year} {2003})}\BibitemShut {NoStop}%
	\bibitem [{\citenamefont {Kienzler}\ \emph {et~al.}(2016)\citenamefont
		{Kienzler}, \citenamefont {Fl\"uhmann}, \citenamefont {Negnevitsky},
		\citenamefont {Lo}, \citenamefont {Marinelli}, \citenamefont {Nadlinger},\
		and\ \citenamefont {Home}}]{kienzler16}%
	\BibitemOpen
	\bibfield  {author} {\bibinfo {author} {\bibfnamefont {D.}~\bibnamefont
			{Kienzler}}, \bibinfo {author} {\bibfnamefont {C.}~\bibnamefont
			{Fl\"uhmann}}, \bibinfo {author} {\bibfnamefont {V.}~\bibnamefont
			{Negnevitsky}}, \bibinfo {author} {\bibfnamefont {H.-Y.}\ \bibnamefont {Lo}},
		\bibinfo {author} {\bibfnamefont {M.}~\bibnamefont {Marinelli}}, \bibinfo
		{author} {\bibfnamefont {D.}~\bibnamefont {Nadlinger}},\ and\ \bibinfo
		{author} {\bibfnamefont {J.~P.}\ \bibnamefont {Home}},\ }\bibfield  {title}
	{\bibinfo {title} {Observation of quantum interference between separated
			mechanical oscillator wave packets},\ }\href
	{https://doi.org/10.1103/PhysRevLett.116.140402} {\bibfield  {journal}
		{\bibinfo  {journal} {Phys. Rev. Lett.}\ }\textbf {\bibinfo {volume} {116}},\
		\bibinfo {pages} {140402} (\bibinfo {year} {2016})}\BibitemShut {NoStop}%
	\bibitem [{\citenamefont {Lv}\ \emph {et~al.}(2018)\citenamefont {Lv},
		\citenamefont {An}, \citenamefont {Liu}, \citenamefont {Zhang}, \citenamefont
		{Pedernales}, \citenamefont {Lamata}, \citenamefont {Solano},\ and\
		\citenamefont {Kim}}]{lv18}%
	\BibitemOpen
	\bibfield  {author} {\bibinfo {author} {\bibfnamefont {D.}~\bibnamefont
			{Lv}}, \bibinfo {author} {\bibfnamefont {S.}~\bibnamefont {An}}, \bibinfo
		{author} {\bibfnamefont {Z.}~\bibnamefont {Liu}}, \bibinfo {author}
		{\bibfnamefont {J.-N.}\ \bibnamefont {Zhang}}, \bibinfo {author}
		{\bibfnamefont {J.~S.}\ \bibnamefont {Pedernales}}, \bibinfo {author}
		{\bibfnamefont {L.}~\bibnamefont {Lamata}}, \bibinfo {author} {\bibfnamefont
			{E.}~\bibnamefont {Solano}},\ and\ \bibinfo {author} {\bibfnamefont
			{K.}~\bibnamefont {Kim}},\ }\bibfield  {title} {\bibinfo {title} {Quantum
			simulation of the quantum rabi model in a trapped ion},\ }\href
	{https://doi.org/10.1103/PhysRevX.8.021027} {\bibfield  {journal} {\bibinfo
			{journal} {Phys. Rev. X}\ }\textbf {\bibinfo {volume} {8}},\ \bibinfo {pages}
		{021027} (\bibinfo {year} {2018})}\BibitemShut {NoStop}%
	\bibitem [{\citenamefont {Bock}\ \emph {et~al.}(2018)\citenamefont {Bock},
		\citenamefont {Eich}, \citenamefont {Kucera}, \citenamefont {Kreis},
		\citenamefont {Lenhard}, \citenamefont {Becher},\ and\ \citenamefont
		{Eschner}}]{bock18}%
	\BibitemOpen
	\bibfield  {author} {\bibinfo {author} {\bibfnamefont {M.}~\bibnamefont
			{Bock}}, \bibinfo {author} {\bibfnamefont {P.}~\bibnamefont {Eich}}, \bibinfo
		{author} {\bibfnamefont {S.}~\bibnamefont {Kucera}}, \bibinfo {author}
		{\bibfnamefont {M.}~\bibnamefont {Kreis}}, \bibinfo {author} {\bibfnamefont
			{A.}~\bibnamefont {Lenhard}}, \bibinfo {author} {\bibfnamefont
			{C.}~\bibnamefont {Becher}},\ and\ \bibinfo {author} {\bibfnamefont
			{J.}~\bibnamefont {Eschner}},\ }\bibfield  {title} {\bibinfo {title}
		{High-fidelity entanglement between a trapped ion and a telecom photon via
			quantum frequency conversion},\ }\href
	{https://doi.org/10.1038/s41467-018-04341-2} {\bibfield  {journal} {\bibinfo
			{journal} {Nature communications}\ }\textbf {\bibinfo {volume} {9}},\
		\bibinfo {pages} {1998} (\bibinfo {year} {2018})}\BibitemShut {NoStop}%
	\bibitem [{\citenamefont {Landsman}\ \emph {et~al.}(2019)\citenamefont
		{Landsman}, \citenamefont {Figgatt}, \citenamefont {Schuster}, \citenamefont
		{Linke}, \citenamefont {Yoshida}, \citenamefont {Yao},\ and\ \citenamefont
		{Monroe}}]{landsman19}%
	\BibitemOpen
	\bibfield  {author} {\bibinfo {author} {\bibfnamefont {K.~A.}\ \bibnamefont
			{Landsman}}, \bibinfo {author} {\bibfnamefont {C.}~\bibnamefont {Figgatt}},
		\bibinfo {author} {\bibfnamefont {T.}~\bibnamefont {Schuster}}, \bibinfo
		{author} {\bibfnamefont {N.~M.}\ \bibnamefont {Linke}}, \bibinfo {author}
		{\bibfnamefont {B.}~\bibnamefont {Yoshida}}, \bibinfo {author} {\bibfnamefont
			{N.~Y.}\ \bibnamefont {Yao}},\ and\ \bibinfo {author} {\bibfnamefont
			{C.}~\bibnamefont {Monroe}},\ }\bibfield  {title} {\bibinfo {title} {Verified
			quantum information scrambling},\ }\href
	{https://doi.org/https://doi.org/10.1038/s41586-019-0952-6} {\bibfield
		{journal} {\bibinfo  {journal} {Nature}\ }\textbf {\bibinfo {volume} {567}},\
		\bibinfo {pages} {61} (\bibinfo {year} {2019})}\BibitemShut {NoStop}%
	\bibitem [{\citenamefont {Monroe}\ \emph {et~al.}(2021)\citenamefont {Monroe},
		\citenamefont {Campbell}, \citenamefont {Duan}, \citenamefont {Gong},
		\citenamefont {Gorshkov}, \citenamefont {Hess}, \citenamefont {Islam},
		\citenamefont {Kim}, \citenamefont {Linke}, \citenamefont {Pagano},
		\citenamefont {Richerme}, \citenamefont {Senko},\ and\ \citenamefont
		{Yao}}]{monroe21}%
	\BibitemOpen
	\bibfield  {author} {\bibinfo {author} {\bibfnamefont {C.}~\bibnamefont
			{Monroe}}, \bibinfo {author} {\bibfnamefont {W.~C.}\ \bibnamefont
			{Campbell}}, \bibinfo {author} {\bibfnamefont {L.-M.}\ \bibnamefont {Duan}},
		\bibinfo {author} {\bibfnamefont {Z.-X.}\ \bibnamefont {Gong}}, \bibinfo
		{author} {\bibfnamefont {A.~V.}\ \bibnamefont {Gorshkov}}, \bibinfo {author}
		{\bibfnamefont {P.~W.}\ \bibnamefont {Hess}}, \bibinfo {author}
		{\bibfnamefont {R.}~\bibnamefont {Islam}}, \bibinfo {author} {\bibfnamefont
			{K.}~\bibnamefont {Kim}}, \bibinfo {author} {\bibfnamefont {N.~M.}\
			\bibnamefont {Linke}}, \bibinfo {author} {\bibfnamefont {G.}~\bibnamefont
			{Pagano}}, \bibinfo {author} {\bibfnamefont {P.}~\bibnamefont {Richerme}},
		\bibinfo {author} {\bibfnamefont {C.}~\bibnamefont {Senko}},\ and\ \bibinfo
		{author} {\bibfnamefont {N.~Y.}\ \bibnamefont {Yao}},\ }\bibfield  {title}
	{\bibinfo {title} {Programmable quantum simulations of spin systems with
			trapped ions},\ }\href {https://doi.org/10.1103/RevModPhys.93.025001}
	{\bibfield  {journal} {\bibinfo  {journal} {Rev. Mod. Phys.}\ }\textbf
		{\bibinfo {volume} {93}},\ \bibinfo {pages} {025001} (\bibinfo {year}
		{2021})}\BibitemShut {NoStop}%
	\bibitem [{\citenamefont {Kokail}\ \emph {et~al.}(2021)\citenamefont {Kokail},
		\citenamefont {van Bijnen}, \citenamefont {Elben}, \citenamefont
		{Vermersch},\ and\ \citenamefont {Zoller}}]{kokail22}%
	\BibitemOpen
	\bibfield  {author} {\bibinfo {author} {\bibfnamefont {C.}~\bibnamefont
			{Kokail}}, \bibinfo {author} {\bibfnamefont {R.}~\bibnamefont {van Bijnen}},
		\bibinfo {author} {\bibfnamefont {A.}~\bibnamefont {Elben}}, \bibinfo
		{author} {\bibfnamefont {B.}~\bibnamefont {Vermersch}},\ and\ \bibinfo
		{author} {\bibfnamefont {P.}~\bibnamefont {Zoller}},\ }\bibfield  {title}
	{\bibinfo {title} {Entanglement hamiltonian tomography in quantum
			simulation},\ }\href {https://doi.org/10.1038/s41567-021-01260-w} {\bibfield
		{journal} {\bibinfo  {journal} {Nature Physics}\ }\textbf {\bibinfo {volume}
			{17}},\ \bibinfo {pages} {936} (\bibinfo {year} {2021})}\BibitemShut
	{NoStop}%
	\bibitem [{\citenamefont {Matsos}\ \emph {et~al.}(2023)\citenamefont {Matsos},
		\citenamefont {Valahu}, \citenamefont {Navickas}, \citenamefont {Rao},
		\citenamefont {Millican}, \citenamefont {Biercuk},\ and\ \citenamefont
		{Tan}}]{matsos23}%
	\BibitemOpen
	\bibfield  {author} {\bibinfo {author} {\bibfnamefont {V.~G.}\ \bibnamefont
			{Matsos}}, \bibinfo {author} {\bibfnamefont {C.~H.}\ \bibnamefont {Valahu}},
		\bibinfo {author} {\bibfnamefont {T.}~\bibnamefont {Navickas}}, \bibinfo
		{author} {\bibfnamefont {A.~D.}\ \bibnamefont {Rao}}, \bibinfo {author}
		{\bibfnamefont {M.~J.}\ \bibnamefont {Millican}}, \bibinfo {author}
		{\bibfnamefont {M.~J.}\ \bibnamefont {Biercuk}},\ and\ \bibinfo {author}
		{\bibfnamefont {T.~R.}\ \bibnamefont {Tan}},\ }\href@noop {} {\bibinfo
		{title} {Robust and deterministic preparation of bosonic logical states in a
			trapped ion}} (\bibinfo {year} {2023}),\ \Eprint
	{https://arxiv.org/abs/2310.15546} {arXiv:2310.15546 [quant-ph]} \BibitemShut
	{NoStop}%
	\bibitem [{\citenamefont {Chaves}\ and\ \citenamefont
		{Davidovich}(2010)}]{chaves10}%
	\BibitemOpen
	\bibfield  {author} {\bibinfo {author} {\bibfnamefont {R.}~\bibnamefont
			{Chaves}}\ and\ \bibinfo {author} {\bibfnamefont {L.}~\bibnamefont
			{Davidovich}},\ }\bibfield  {title} {\bibinfo {title} {Robustness of
			entanglement as a resource},\ }\href
	{https://doi.org/10.1103/PhysRevA.82.052308} {\bibfield  {journal} {\bibinfo
			{journal} {Phys. Rev. A}\ }\textbf {\bibinfo {volume} {82}},\ \bibinfo
		{pages} {052308} (\bibinfo {year} {2010})}\BibitemShut {NoStop}%
	\bibitem [{\citenamefont {Chitambar}\ and\ \citenamefont
		{Gour}(2019)}]{chitambar19}%
	\BibitemOpen
	\bibfield  {author} {\bibinfo {author} {\bibfnamefont {E.}~\bibnamefont
			{Chitambar}}\ and\ \bibinfo {author} {\bibfnamefont {G.}~\bibnamefont
			{Gour}},\ }\bibfield  {title} {\bibinfo {title} {Quantum resource theories},\
	}\href {https://doi.org/10.1103/RevModPhys.91.025001} {\bibfield  {journal}
		{\bibinfo  {journal} {Rev. Mod. Phys.}\ }\textbf {\bibinfo {volume} {91}},\
		\bibinfo {pages} {025001} (\bibinfo {year} {2019})}\BibitemShut {NoStop}%
	\bibitem [{\citenamefont {Roszak}\ and\ \citenamefont
		{Cywi{\'{n}}ski}(2015)}]{roszak15b}%
	\BibitemOpen
	\bibfield  {author} {\bibinfo {author} {\bibfnamefont {K.}~\bibnamefont
			{Roszak}}\ and\ \bibinfo {author} {\bibfnamefont {{\L}.}~\bibnamefont
			{Cywi{\'{n}}ski}},\ }\bibfield  {title} {\bibinfo {title} {The relation
			between the quantum discord and quantum teleportation: The physical
			interpretation of the transition point between different quantum discord
			decay regimes},\ }\href {https://doi.org/10.1209/0295-5075/112/10002}
	{\bibfield  {journal} {\bibinfo  {journal} {{EPL} (Europhysics Letters)}\
		}\textbf {\bibinfo {volume} {112}},\ \bibinfo {pages} {10002} (\bibinfo
		{year} {2015})}\BibitemShut {NoStop}%
	\bibitem [{\citenamefont {Rzepkowski}\ and\ \citenamefont
		{Roszak}(2023)}]{rzepkowski23}%
	\BibitemOpen
	\bibfield  {author} {\bibinfo {author} {\bibfnamefont {B.}~\bibnamefont
			{Rzepkowski}}\ and\ \bibinfo {author} {\bibfnamefont {K.}~\bibnamefont
			{Roszak}},\ }\bibfield  {title} {\bibinfo {title} {Signature of quantumness
			in pure decoherence control},\ }\href
	{https://doi.org/10.1103/PhysRevA.108.012412} {\bibfield  {journal} {\bibinfo
			{journal} {Phys. Rev. A}\ }\textbf {\bibinfo {volume} {108}},\ \bibinfo
		{pages} {012412} (\bibinfo {year} {2023})}\BibitemShut {NoStop}%
	\bibitem [{\citenamefont {Flühmann}(2019)}]{fluhmann19}%
	\BibitemOpen
	\bibfield  {author} {\bibinfo {author} {\bibfnamefont {C.}~\bibnamefont
			{Flühmann}},\ }{\selectlanguage {en}\emph {\bibinfo {title} {Encoding a
				qubit in the motion of a trapped ion using superpositions of displaced
				squeezed states}}},\ \href {https://doi.org/10.3929/ethz-b-000355836}
	{\bibinfo {type} {Doctoral thesis}},\ \bibinfo  {school} {ETH Zurich},
	\bibinfo {address} {Zurich} (\bibinfo {year} {2019})\BibitemShut {NoStop}%
\end{thebibliography}
\end{document}